\documentclass[twocolumn,english,journal]{IEEEtran}

\usepackage{graphicx}
\usepackage{color}
\usepackage{placeins}
\usepackage{float}
\usepackage{tabularx,colortbl}
\usepackage{amssymb}
\usepackage{amsthm}
\usepackage{cite}
\usepackage{amsmath}
\usepackage{caption}
\usepackage{subcaption}
\usepackage{url}

\def\vectorize{\mathrm{vec}}

\def\kron{\otimes}
\def\tr{\mathrm{tr}}

\def\vectorize{\mathrm{vec}}

\def\Htran{\mbox{\tiny $\mathrm{H}$}}
\def\Ttran{\mbox{\tiny $\mathrm{T}$}}

\newcommand{\fracSum}[1]{{\underset{{#1}}{\sum}}}

\newcommand{\fracSumtwo}[2]{\overset{#2}{\underset{#1}{\sum}}}
\newcommand{\vect}[1]{\mathbf{#1}}

\theoremstyle{plain}
\newtheorem{remark}{Remark}

\newtheorem{theorem}{Theorem}
\newtheorem{corollary}{Corollary}
\newtheorem{lemma}{Lemma}

\begin{document}

\title{\huge{Massive MIMO for Maximal Spectral Efficiency: \\ How Many Users and Pilots Should Be Allocated?}}
\author{Emil Bj{\"o}rnson, \emph{Member, IEEE}, Erik G.~Larsson, \emph{Senior Member, IEEE}, and
M{\'e}rouane Debbah, \emph{Fellow, IEEE}
\thanks{E. Bj\"ornson and E.~G. Larsson are with Department of Electrical Engineering (ISY), Link\"{o}ping University, Link\"{o}ping, Sweden (\{emil.bjornson,erik.larsson\}@liu.se). M. Debbah is with CentraleSup{\'e}lec, Gif-sur-Yvette, France (merouane.debbah@supelec.fr) and with the Mathematical and Algorithmic Sciences Lab, Huawei, Paris, France.
\newline \indent This research has received funding from the EU FP7 under ICT-619086 (MAMMOET), from ELLIIT, the Swedish Research Council (VR), and the ERC Grant 305123 MORE.
\newline \indent Part of the material in this paper was presented at the IEEE Global Conference on Signal and Information Processing (GlobalSIP), Atlanta, Georgia, December 3-5, 2014.}}

\IEEEoverridecommandlockouts

\maketitle

\begin{abstract}
Massive MIMO is a promising technique to increase the spectral efficiency (SE) of cellular networks, by deploying antenna arrays with hundreds or thousands of active elements at the base stations and performing coherent transceiver processing. A common rule-of-thumb is that these systems should have an order of magnitude more antennas, $M$, than scheduled users, $K$, because the users' channels are likely to be near-orthogonal when $M/K > 10$. However, it has not been proved that this rule-of-thumb actually maximizes the SE. In this paper, we analyze how the optimal number of scheduled users, $K^\star$, depends on $M$ and other system parameters. To this end, new SE expressions are derived to enable efficient system-level analysis with power control, arbitrary pilot reuse, and random user locations. The value of $K^\star$ in the large-$M$ regime is derived in closed form, while simulations are used to show what happens at finite $M$, in different interference scenarios, with different pilot reuse factors, and for different processing schemes. Up to half the coherence block should be dedicated to pilots and the optimal $M/K$ is less than 10 in many cases of practical relevance. Interestingly, $K^\star$ depends strongly on the processing scheme and hence it is unfair to compare different schemes using the same $K$.
\end{abstract}


\IEEEpeerreviewmaketitle

\section{Introduction}

Cellular communication networks are continuously evolving to keep up with the rapidly increasing demand for wireless data services. Higher area throughput (in bit/s per $\textrm{km}^2$) has traditionally been achieved by a combination of three multiplicative factors \cite{Nokia2011}: more frequency spectrum (Hz), higher cell density (more cells per $\textrm{km}^2$), and higher spectral efficiency (bit/s/Hz/cell). This paper considers the latter and especially the massive multiple-input multiple-output (MIMO) concept, proposed in \cite{Marzetta2010a}, which has been identified as the key to increase the spectral efficiency (SE) by orders of magnitude over contemporary systems \cite{Baldemair2013a,Boccardi2014a,Larsson2014a}.

The massive MIMO concept is based on equipping base stations (BSs) with hundreds or thousands of antenna elements which, unlike conventional cellular technology, are operated in a coherent fashion. This can provide unprecedented array gains and a spatial resolution that allows for multi-user MIMO communication to tens or hundreds of user equipments (UEs) per cell, while maintaining robustness to inter-user interference. The research on massive MIMO has so far focused on establishing the fundamental physical (PHY) layer properties; in particular, that the acquisition of channel state information (CSI) is limited by the channel coherence block (i.e., the fact that channel responses are only static in limited time/frequency blocks) and how this impacts the SEs and the ability to mitigate inter-cell interference \cite{Marzetta2010a,Jose2011b,Hoydis2013a}. In addition, the aggressive multiplexing in massive MIMO has been shown to provide major improvements in the overall energy efficiency \cite{Ngo2013a,Bjornson2015a,Ha2013a,Yang2013a}, while \cite{Bjornson2014a,Bjornson2015b,Pitarokoilis2015a} have shown that the hardware impairments of practical transceivers have smaller impact on massive MIMO than contemporary systems. In contrast, the research community has only briefly touched on the resource allocation problems in the media access control (MAC) layer (e.g., user scheduling)---although the truly achievable SEs can only be understood if the PHY and MAC layers are jointly optimized.

The importance of resource allocation for massive MIMO was described in \cite{Huh2012a}, where initial guidelines were given. A main insight is that the limited number of orthogonal pilot sequences needs to be allocated intelligently among the UEs to reduce interference, which can be done by capitalizing on pathloss differences \cite{Li2012a,Mueller2013a} and spatial correlation \cite{Huh2012a,Yin2013a,Li2013a}.

In this paper, we consider a related resource allocation question: \emph{how many UEs should be scheduled per cell to maximize the spectral efficiency?} This question has, to the best of our knowledge, not been answered for multi-cell systems.\footnote{A few results for single-cell systems are available in the literature; for example, in \cite{Ngo2013a}.} We show how the coherence block length, number of antennas, pilot allocation, hardware impairments, and other system parameters determine the answer. To this end, we derive new SE expressions which are valid for both uplink (UL) and  downlink (DL) transmission, with random user locations and power control that yields uniform UE performance. We consider both conventional linear processing schemes such as maximum ratio (MR) combining/transmission  and zero-forcing (ZF), and a new full-pilot zero-forcing (P-ZF) scheme that actively suppresses inter-cell interference in a fully distributed coordinated beamforming fashion. {The following are the main contributions of each section:}

\begin{itemize}
\item Section \ref{sec:system-model} presents the UL/DL massive MIMO system model, where the unique features are the power control and random UE locations.

\item Section \ref{sec:average-SE} provides new analytic results for channel estimation with arbitrary pilot signals and new tractable SE expressions for the UL and DL with random UE locations and power control. MR and ZF processing are considered, as well as the new P-ZF scheme.

\item Section \ref{sec:optimization-hexagonal} provides extensive simulation results on the maximal SE, where the impact of all system parameters are explained. The expected massive MIMO gains are illustrated.

\item Section \ref{sec:hardware-impairments} extends the previous results to systems with hardware impairments.

\item Finally, Section \ref{sec:conclusion} summarizes the main results and insights obtained in the paper.

\end{itemize}

\section{System Model}
\label{sec:system-model}

\begin{figure}[!t]
\begin{center} 
\includegraphics[width=.8\columnwidth]{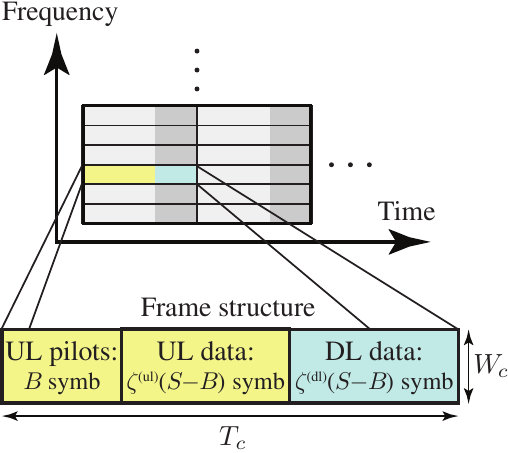}
\end{center}\vskip-3mm
\caption{The transmission is divided into frames of $S = T_c W_c$ symbols, whereof $B$ symbols are dedicated to pilot transmission. The remaining $S-B$ symbols are used for payload data, where $\zeta^{\rm{(ul)}}$ and $\zeta^{\rm{(dl)}}$ are respectively the fractions of UL and DL transmission.} \label{figure_protocol} \vskip-3mm
\end{figure}

{ We consider a cellular network where payload data is transmitted with universal time and frequency reuse. Each cell is assigned an index in the set $\mathcal{L}$, where the cardinality $|\mathcal{L}|$ is the number of cells. The BS in each cell is equipped with an array of $M$ antennas and communicates with $K$ single-antenna UEs at the time, out of a set of $K_{\max}$ UEs. We are interested in massive MIMO topologies where $M$ and $K_{\max}$ are large and fixed, while $K$ is a design parameter and all UEs have unlimited demand for data. The subset of active UEs changes over time, thus the name UE $k \in \{1,\ldots,K\}$ in cell $l\in \mathcal{L}$ is given to different UEs at different times. The geographical position $\vect{z}_{lk} \in \mathbb{R}^2$ of  UE $k$ in cell $l$ is therefore an ergodic random variable with a cell-specific distribution. This model is used to study the average performance for a random rather than fixed set of interfering UEs.} The time-frequency resources are divided into frames consisting of $T_c$ seconds and $W_c$ Hz, as illustrated in Fig.~\ref{figure_protocol}.\footnote{ This paper concentrates on frames that carry user-specific signals, in particular, payload data and pilots. From time to time, the network also needs special frames to transmit cell-specific control and system information and to enable random access. The design of these control frames is outside the scope of this paper, but some initial results are found in \cite{Karlsson2014a}.} This leaves room for $S = T_c W_c$ transmission symbols per frame. We assume that the frame dimensions are such that $T_c$ is smaller or equal to the coherence time of all UEs, while $W_c$ is smaller or equal to the coherence bandwidth of all UEs. Hence, all the channels are static within the frame; $\vect{h}_{jlk} \in \mathbb{C}^N$ denotes the channel response between BS $j$ and UE $k$ in cell $l$ in a given frame.
These channel responses are drawn as realizations from zero-mean circularly symmetric complex Gaussian distributions:
\begin{equation} \label{eq:channel-distribution}
\vect{h}_{jlk} \sim \mathcal{CN}\Big(\vect{0},d_j(\vect{z}_{lk}) \vect{I}_M \Big),
\end{equation}
where $\vect{I}_M$ is the $M \times M$ identity matrix. {This is a theoretical model for non-line-of-sight propagation that is known to give representative results with both few and many BS antennas (see recent channel measurements reported in \cite{Gao2015a}).} The deterministic function $d_j(\vect{z})$ gives the variance of the channel attenuation from BS $j$ to any UE position $\vect{z}$. {The value of $d_j(\vect{z}_{lk})$ varies slowly over time and frequency, thus we assume that the value is known at BS $j$ for all $l$ and $k$ and that each UE knows its value to its serving BS. The exact UE positions $\vect{z}_{lk}$ are unknown.}

We consider the time-division duplex (TDD) protocol shown in Fig.~\ref{figure_protocol}, where $B \geq 1$ out of the $S$ symbols in each frame are reserved for UL pilot signaling. There is no DL pilot signaling and no feedback of CSI, because the BSs can process both UL and DL signals using the UL channel measurements due to the channel reciprocity in TDD systems. The remaining $S-B$ symbols are allocated for payload data and are split between UL and DL transmission. We let $\zeta^{\rm{(ul)}}$ and $\zeta^{\rm{(dl)}}$ denote the fixed fractions allocated for UL and DL, respectively. These fractions can be selected arbitrarily, subject to the constraint $\zeta^{\rm{(ul)}} + \zeta^{\rm{(dl)}} = 1$ and that $\zeta^{\rm{(ul)}} (S-B)$ and $\zeta^{\rm{(dl)}} (S-B)$ are positive integers. Below, we define the system models for the UL and DL.

The BSs are not exchanging any short-term information in this work, but we will see how the pilot allocation and transmission processing can be coordinated in a distributed fashion.

\subsection{Uplink}

The received UL signal $\vect{y}_j \in \mathbb{C}^{M}$ at BS $j$ in a frame is modeled, similar to \cite{Hoydis2013a} and \cite{Ngo2013a}, as
\begin{equation} \label{eq:system-model}
\vect{y}_j = \sum_{l \in \mathcal{L}} \sum_{k=1}^{K} \sqrt{p_{lk}} \vect{h}_{jlk} x_{lk} + \vect{n}_{j}
\end{equation}
where $x_{lk} \in \mathbb{C}$ is the symbol transmitted by UE $k$ in cell $l$. This signal is normalized as $\mathbb{E}\{ | x_{lk} |^2 \} = 1$, while the corresponding UL transmit power is defined by $p_{lk} \geq 0$. The additive noise $\vect{n}_{j}  \in \mathbb{C}^{M}$ is modeled as  $\vect{n}_{j} \sim \mathcal{CN}(\vect{0},\sigma^2 \vect{I}_M)$, where $\sigma^2$ is the noise variance.

Contrary to most previous works on massive MIMO, which assume fixed UL power, we consider statistics-aware power control\footnote{Channel-aware power control was considered in \cite{Guo2014a} and \cite{Bjornson2015a}, but it requires a rapid feedback mechanism where UEs are provided with instantaneous CSI. Since the small-scale fading average out in massive MIMO systems \cite{Marzetta2010a}, statistical power control policies are expected to be almost equal to channel-aware policies \cite{Yang2014a}, but are considerably easier to implement.}; the symbols from UE $k$ in cell $l$ have the transmit power $p_{lk} = \frac{\rho}{d_l(\vect{z}_{lk})}$, where $\rho>0$ is a design parameter.\footnote{ The parameter $\rho$ needs to be selected such that UEs at the cell edge do not use more transmit power than their amplifiers can handle or the spectrum regulations allow. This is not a critical limitation in massive MIMO since high SEs are provided also at low SNRs (see Fig.~\ref{figure_journal_SNR}), but it might be necessary to occasionally drop severely shadowed UEs from service.} This power-control policy inverts the average channel attenuation $d_l(\vect{z}_{lk})$ and has the merit of making the average effective channel gain the same for all UEs: $\mathbb{E}\{ p_{lk} \|\vect{h}_{llk}\|^2 \} = M \rho$. Hence, this policy guarantees a uniform user experience, saves valuable energy at UEs, and avoids near-far blockage where weak signals drown in stronger signals due to the finite dynamic range of analog-to-digital converters (ADCs).

\vspace{-4mm}

\subsection{Downlink}

Building on the UL/DL channel reciprocity in calibrated TDD systems, the received DL signal $z_{jk} \in \mathbb{C}$  at UE $k$ in cell $j$ in a frame is modeled as
\begin{equation} \label{eq:system-model-DL}
z_{jk} = \sum_{l \in \mathcal{L}} \sum_{m=1}^{K} \vect{h}_{ljk}^{\Ttran} \vect{w}_{lm} s_{lm} + \eta_{jk}
\end{equation}
where $(\cdot)^{\Ttran}$ denotes transpose, $s_{lm}$ is the symbol intended for UE $m$ in cell $l$, $\vect{w}_{lm} \in \mathbb{C}^{M}$ is the corresponding precoding vector, and $\|\vect{w}_{lm} \|^2$ is the allocated DL transmit power. Any power control can be considered in the DL since the BS has access to the estimated  CSI. We show later how to select the transmit power to achieve the same SEs in the DL as in the UL. The additive noise at UE $k$ in cell $j$ is modeled as $\eta_{jk} \sim \mathcal{CN}(0,\sigma^2)$, with the same variance as in the UL.\footnote{The noise variance is conventionally lower in the UL, due to better hardware characteristics at the BS, but since massive MIMO has an inherent robustness to noise amplification \cite{Bjornson2015b} it is possible to use handset-like hardware at the BSs. In any case, any disparity in noise power between the UL and DL can be absorbed into the transmit powers without loss of generality.}

\begin{remark}[Synchronization Issues] 
The UL/DL system models in \eqref{eq:system-model} and \eqref{eq:system-model-DL} assume perfect synchronization across all cells, as commonly done in the massive MIMO literature; cf.~\cite{Marzetta2010a,Jose2011b,Hoydis2013a,Ngo2013a,Huh2012a}. Local synchronization is achievable, for example, using the cyclic prefix in OFDM-based systems, but network-wide synchronization is probably infeasible over large coverage areas. The processing techniques analyzed in this paper can thus be used to suppress the strong interference between the closest tiers of neighboring cells, while the interference from distant cells is asynchronously received and practically insuppressible. We expect that the simplified synchronization modeling used here and elsewhere has negligible impact on the system performance, since the insuppressible distant interferers are weak as compared to (partially suppressed) interference from neighboring cells.
\end{remark}

\section{Average Per-Cell Spectral Efficiency}
\label{sec:average-SE}

In this section, we derive and analyze the SE for multi-cell systems with random UE positions.

\subsection{Pilot-Based Channel Estimation}

BS $j$ can use its multitude of antennas for coherent receive combining in the UL and transmit precoding in the DL, which can adaptively amplify desired signals and suppress interfering signals. This requires, however, some knowledge of the UEs' channels; for example, $\sqrt{p_{lk}} \vect{h}_{jlk}$ in the UL, for all $l$ and $k$. Such CSI is typically acquired by pilot signaling, where the UEs send known signals in a predefined manner. Accurate CSI acquisition is a challenging task in multi-cell systems, where the transmission resources are reused across cells, because the pilot signals are inevitably affected by inter-cell interference. This so-called \emph{pilot contamination} limits the quality of the acquired CSI and the ability to reject inter-cell interference (unless intricate subspace methods can be used for decontamination, as suggested in \cite{Mueller2013a}).

The impact of pilot contamination is usually studied under the assumption that exactly the same pilot signals are used in all cells. In contrast, this section derives the main properties of massive MIMO systems (with power control) for arbitrary pilot reuse, where each cell might only use a subset of the pilots. As shown in Fig.~\ref{figure_protocol}, the pilot signals are assumed to span $B$ symbols of each frame, where $1 \leq B \leq S$.\footnote{The pilot signals need not be synchronized across the cells as assumed herein, but there is little to gain from shifting the pilot signals and UL payload data signals between cells; this leads to a mix of deterministic pilots and stochastic data signals at each symbol transmission, but the average pilot contamination will not change in any substantial way \cite[Remark 5]{Ngo2013a}. The new full-pilot interference suppression concepts proposed in this paper are also harder to implement in such cases.} Each pilot signal can be represented by a deterministic vector $\vect{v} \in \mathbb{C}^B$ and the fixed per-symbol power implies that all entries have unit magnitude: $| [ \vect{v} ]_s | = 1$, where $[\cdot]_s$ denotes the $s$th element for $s\in \{ 1, \ldots,B\}$. We assume that all pilot signals originate from a fixed \emph{pilot book} $\mathcal{V}$, defined as
\begin{equation}
\mathcal{V} = \{ \vect{v}_1,\ldots,\vect{v}_B\} \quad \text{where} \quad \vect{v}_{b_1}^{\Htran} \vect{v}_{b_2} = \begin{cases} B, & b_1 = b_2,\\ 0, & b_1 \neq b_2, \end{cases}
\end{equation}
where $(\cdot)^{\Htran}$ denotes the conjugate transpose. Hence, the $B$ pilot signals form an orthogonal basis and can, for example, be the columns of a discrete Fourier transform (DFT) matrix \cite{Biguesh2004a}.

The pilot signal transmitted by UE $k$ in cell $l$ is denoted by $\vect{v}_{i_{lk}}$, where $i_{lk} \in \{1,\ldots,B\}$ is the index in the pilot book. By transmitting these pilot signals over $B$ symbols in the UL system model of \eqref{eq:system-model-pilots}, the collective received UL signal at BS $j$ is denoted as $\vect{Y}_j \in \mathbb{C}^{M \times B}$ and given by
\begin{equation} \label{eq:system-model-pilots}
\vect{Y}_j = \sum_{l \in \mathcal{L}} \sum_{k=1}^{K} \sqrt{p_{lk}} \vect{h}_{jlk} \vect{v}_{i_{lk}}^{\Ttran} + \vect{N}_{j},
\end{equation}
where $\vect{N}_{j} \in \mathbb{C}^{M \times B}$ contains the additive noise at the receiver during the pilot signaling.

\begin{figure*}[!t]
% ensure that we have normalsize text
\normalsize

\begin{equation} \label{eq:SINR-value} \tag{12}
\mathrm{SINR}_{jk}^{\rm{(ul)}} =  \frac{ p_{jk} | \mathbb{E}_{\{\vect{h}\}}\{ \vect{g}_{jk}^{\Htran} \vect{h}_{jjk} \} |^2 }{ \fracSum{l \in \mathcal{L}} \fracSumtwo{m=1}{K} p_{lm}  \mathbb{E}_{\{\vect{h}\}} \{ |\vect{g}_{jk}^{\Htran} \vect{h}_{jlm} |^2  \} - p_{jk} | \mathbb{E}_{\{\vect{h}\}} \{ \vect{g}_{jk}^{\Htran} \vect{h}_{jjk} \} |^2  + \sigma^2  \mathbb{E}_{\{\vect{h}\}} \{ \| \vect{g}_{jk} \|^2\}  }.
\end{equation}
% Restore the current equation number.
% IEEE uses as a separator
\hrulefill
\end{figure*}

The following lemma derives the minimum mean-squared error (MMSE) estimator of the effective power-controlled UL channels, which are defined as $\vect{h}_{jlk}^{\mathrm{eff}} = \sqrt{p_{lk}} \vect{h}_{jlk}$.

\begin{lemma} \label{lemma:LMMSE-estimation}
The MMSE estimate at BS $j$ of the effective power-controlled UL channel $\vect{h}_{jlk}^{\mathrm{eff}}$, for any UE $k\in \{1,\ldots,K\}$ in any cell $l \in \mathcal{L}$, is
\begin{equation} \label{eq:LMMSE-estimator}
  \hat{\vect{h}}_{jlk}^{\mathrm{eff}} = \frac{d_j(\vect{z}_{lk}) }{ d_l(\vect{z}_{lk})} \vect{Y}_j (\boldsymbol{\Psi}^{\Ttran}_j)^{-1} \vect{v}_{i_{lk}}^{*}
\end{equation}
where $(\cdot)^*$ denotes the complex conjugate and the normalized covariance matrix
$\boldsymbol{\Psi}_j \in \mathbb{C}^{B \times B}$ of the received signal is
\begin{align}
\boldsymbol{\Psi}_j &= \sum_{ \ell \in \mathcal{L}} \sum_{m=1}^{K} \frac{d_j(\vect{z}_{\ell m}) }{ d_{\ell}(\vect{z}_{\ell m})}  \vect{v}_{i_{\ell m}} \vect{v}_{i_{\ell m}}^{\Htran} + \frac{\sigma^2}{\rho}  \vect{I}_B \label{eq:Cerror-def}.
\end{align}
The estimation error covariance matrix $\vect{C}_{jlk} \in \mathbb{C}^{M \times M}$ is given by
\begin{equation} \label{eq:LMMSE-error-cov}
\begin{split}
&\vect{C}_{jlk} = \mathbb{E}\left\{ ( \vect{h}_{jlk}^{\mathrm{eff}} -\hat{\vect{h}}_{jlk}^{\mathrm{eff}} )( \vect{h}_{jlk}^{\mathrm{eff}} - \hat{\vect{h}}_{jlk}^{\mathrm{eff}} )^{\Htran}   \right\} \\ &= \rho \frac{d_j(\vect{z}_{lk}) }{ d_l(\vect{z}_{lk})} \left( 1 - \frac{\frac{d_j(\vect{z}_{lk}) }{ d_l(\vect{z}_{lk})} B}{\sum_{\ell \in \mathcal{L}} \sum_{m=1}^{K} \frac{d_j(\vect{z}_{\ell m}) }{ d_{\ell}(\vect{z}_{\ell m})} \vect{v}_{i_{lk}}^{\Htran} \vect{v}_{i_{\ell m}} + \frac{\sigma^2}{\rho}} \right) \vect{I}_M
\end{split}
\end{equation}
and the mean-squared error (MSE) is $\mathrm{MSE}_{jlk} = \tr( \vect{C}_{jlk})$.
\end{lemma}
\begin{IEEEproof}
The proof is given in the appendix.
\end{IEEEproof}

There are two important differences between Lemma \ref{lemma:LMMSE-estimation} and the channel estimators that are conventionally used in the massive MIMO literature: 1) we estimate the effective channels including the UL power control; and 2) the MMSE estimator supports arbitrary pilot allocation.

The covariance matrix in \eqref{eq:LMMSE-error-cov} reveals the causes of estimation errors; it depends on the inverse signal-to-noise ratio (SNR), $\sigma^2/\rho$, and on which UEs that use the same pilot signal (i.e., which of the products $\vect{v}_{i_{lk}}^{\Htran} \vect{v}_{i_{\ell m}} $ that are non-zero). The ratio $d_j(\vect{z}_{\ell m}) / d_{\ell}(\vect{z}_{\ell m})$ describes the relative strength of the interference received at BS $j$ from UE $m$ in cell $\ell$; it is almost one for cell-edge UEs of neighboring cells, while it is almost zero when cell $\ell$ is very distant from BS $j$.

Although Lemma \ref{lemma:LMMSE-estimation} allows for estimation of all channel vectors in the whole cellular network, each BS can only resolve $B$ different spatial dimensions since there are only $B$ orthogonal pilot signals. To show this explicitly, we define the $M \times B$ matrix
\begin{equation}
\widehat{\vect{H}}_{\mathcal{V},j} = \vect{Y}_j (\boldsymbol{\Psi}^{\Ttran}_j)^{-1}  \left[  \vect{v}_1^{*} \, \ldots, \vect{v}_B^{*}  \right]
\end{equation}
using each of the $B$ pilot signals from $\mathcal{V}$. The channel estimate in \eqref{eq:LMMSE-estimator} for UE $k$ in cell $l$, which uses the pilot $\vect{v}_{i_{lk}}$, is parallel to the $i_{lk}$th column of $\widehat{\vect{H}}_{\mathcal{V},j}$; more precisely, we have
\begin{equation}
\hat{\vect{h}}_{jlk}^{\mathrm{eff}} = \frac{d_j(\vect{z}_{lk}) }{ d_l(\vect{z}_{lk})} \widehat{\vect{H}}_{\mathcal{V},j} \vect{e}_{i_{lk}}
\end{equation}
where $\vect{e}_{i}$ denotes the $i$th column of the identity matrix $\vect{I}_B$. This is the essence of pilot contamination; BSs cannot tell apart UEs that use the same pilot signal and cannot reject the corresponding interference since the estimated channels are parallel.  In some cases (e.g., for slow changes in the user scheduling and high spatial channel correlation), statistical prior knowledge can be used to partially separate the UEs \cite{Yin2013a}, {but this possibility is not considered herein since we want to develop methods to suppress pilot contamination that can be utilized in any propagation environment.}

\begin{remark}[Mobility and Pilot Sharing]
Each UE might have a different dimension of its coherence block, defined by some coherence time $\tilde{T}_c$ and coherence bandwidth $\tilde{W}_c$, depending on the propagation environment and the UE's mobility. Suppose that $\tilde{T}_c = a T_c$ and $\tilde{W}_c = b W_c$ for a certain UE, where $a \geq 1$ and $b \geq 1$ since the frame structure was defined to fit into the coherence block of all UEs. Then, $\tau = \lfloor a \rfloor \lfloor b \rfloor$ is the total number of frames that fits into the coherence block of this particular UE, where $\lfloor \cdot \rfloor$ stands for truncation. If $\tau>1$, there is no need to send pilots in every frame; it is sufficient with $1/\tau$ of the frames. Hence, multiple UEs with $\tau > 1$ can share a pilot signal, without disturbing one another, by using it in different frames.
\end{remark}

\subsection{Achievable UL Spectral Efficiencies}

The channel estimates in Lemma \ref{lemma:LMMSE-estimation} enable each BS to (semi-)coherently detect the data signals from its UEs. In particular, we assume that BS $j$ applies a linear receive combining vector $\vect{g}_{jk} \in \mathbb{C}^{M}$ to the received signal, as $\vect{g}_{jk}^{\Htran} \vect{y}_j$, to amplify the signal from its $k$th UE and reject interference from other UEs in the spatial domain. {We want to derive the ergodic achievable SE for any UE, where codewords span over both the Rayleigh fading and random locations of the interfering UEs---specific UE distributions are considered in Section \ref{sec:optimization-hexagonal}. For notational convenience, we assume that $\beta = \frac{B}{K}$ is an integer that we refer to as the \emph{pilot reuse factor}. The cells in $\mathcal{L}$ are divided into $\beta \geq 1$ disjoint subsets such that the same $K$ pilot sequences are used within a set, while different pilots are used in different sets. We refer to this as \emph{non-universal pilot reuse}. An explicit example is provided in Section \ref{sec:optimization-hexagonal} for hexagonal cells, while the result in this section holds for any network topology.} The following lemma shows how the SE depends on the receive combining, for Gaussian codebooks where $x_{jk} \sim \mathcal{CN}(0,1)$.

\begin{lemma} \label{lemma:SE}
In the UL, an ergodic achievable SE of {an arbitrary} UE $k$ in cell $j$ is
\begin{equation}
\zeta^{\rm{(ul)}} \left( 1-\frac{B}{S} \right) \mathbb{E}_{\{\vect{z}\}} \left\{ \log_2(1+ \mathrm{SINR}_{jk}^{\rm{(ul)}}) \right\} \quad \text{[bit/s/Hz]}
\end{equation}
where the effective signal-to-interference-and-noise ratio (SINR), $\mathrm{SINR}_{jk}^{\rm{(ul)}}$, is given in \eqref{eq:SINR-value} at the top of the page. The expectations $\mathbb{E}_{\{\vect{z}\}} \{ \cdot \}$  and $\mathbb{E}_{\{\vect{h}\}} \{ \cdot \}$ are with respect to UE positions and channel realizations, respectively. \setcounter{equation}{12}
\end{lemma}
\begin{IEEEproof}
{ By coding over variations in the channel realizations $\{\vect{h}\}$ and positions $\{\vect{z}\}$  of the interfering UEs, an achievable SE is given by $\mathbb{E}_{\{\vect{z},\vect{h}\}} \{ \mathcal{I}(x_{lk},\vect{y}_j) \}$, where $\mathcal{I}(x_{lk},\vect{y}_j)$ is the mutual information between the transmitted and received signal in \eqref{eq:system-model} for fixed channel realizations and UE positions. The lemma follows from computing a lower bound on $\mathcal{I}(x_{lk},\vect{y}_j)$, similar to \cite{Medard2000a,Jose2011b,Hoydis2013a,Bjornson2015b,Hassibi2003a}, by making three limiting assumptions: 1) a Gaussian codebook is used;  2) the signal component received over the effective channel mean $\mathbb{E}_{\{\vect{h}\}}\{ \vect{g}_{jk}^{\Htran} \vect{h}_{jjk} \}$ is the only desired signal, while the interference and the signal component over the remaining uncorrelated channel $\vect{g}_{jk}^{\Htran} \vect{h}_{jjk} - \mathbb{E}_{\{\vect{h}\}}\{ \vect{g}_{jk}^{\Htran} \vect{h}_{jjk} \}$ are treated as noise (i.e., not exploited in the decoding); and 3) the noise is taken as worst-case Gaussian distributed in the decoding, leading to a further lower bound on the mutual information.}
\end{IEEEproof}

{The ergodic achievable SE in Lemma \ref{lemma:SE}, for any UE in cell $j$, is a lower bound on the ergodic capacity, which is unknown for general multi-cell networks. Similar bounds are found in \cite{Jose2011b,Hoydis2013a,Ngo2013a} and the bounding technique interacts with the Rayleigh fading, which is why its expectations end up inside the logarithm while the user positions are averaged at the outside. To compute these expectations we need to specify the receive combining.} The combining schemes for massive MIMO can have either \emph{passive} or \emph{active} interference rejection. The canonical example of passive rejection is maximum ratio (MR) combining, defined as
\begin{equation} \label{eq:def-MR}
\vect{g}_{jk}^{\mathrm{MR}} = \widehat{\vect{H}}_{\mathcal{V},j} \vect{e}_{i_{jk}} = \hat{\vect{h}}_{jjk}^{\mathrm{eff}},
\end{equation}
which maximizes the gain of the desired signal and relies on that interfering signals are rejected automatically since the co-user channels are quasi-orthogonal to $\hat{\vect{h}}_{jjk}^{\mathrm{eff}}$ when $M$ is large.\footnote{With \emph{quasi}-orthogonality we mean that two vectors $\vect{a},\vect{b} \in \mathbb{C}^{M}$ satisfy $\frac{\vect{a}^{\Htran} \vect{b}}{M} \rightarrow 0$ as $M \rightarrow \infty$, although $\vect{a}^{\Htran} \vect{b}$ will not converge to zero and might even go to infinity, e.g., proportionally to $\sqrt{M}$ as with Rayleigh fading channel vectors.}

In contrast, active rejection is achieved by making the receive combining as orthogonal to the interfering channels as possible.
This is conventionally achieved by zero-forcing (ZF) combining, where the combining is selected to orthogonalize the $K$ intra-cell channels:
\begin{equation} \label{eq:def-ZF}
\vect{g}_{jk}^{\mathrm{ZF}} = \widehat{\vect{H}}_{\mathcal{V},j} \vect{E}_j \big( \vect{E}_j^{\Htran} \widehat{\vect{H}}_{\mathcal{V},j}^{\Htran} \widehat{\vect{H}}_{\mathcal{V},j} \vect{E}_j \big)^{-1} \vect{e}_{i_{jk}},
\end{equation}
where $\vect{E}_j = [\vect{e}_{i_{j1}} \, \ldots \vect{e}_{i_{jK}}] \in \mathbb{C}^{B \times K}$ and all the UEs in cell $j$ are required to use different pilots.

The next theorem provides closed-form expressions for the per-cell SEs with MR and ZF.

\begin{theorem} \label{theorem-achievable-rate-conventional}
 Let $\mathcal{L}_j(\beta) \subset \mathcal{L}$ be the subset of cells that uses the same pilots as cell $j$. In the UL, an achievable SE in cell $j$ is
\begin{equation} \label{eq:UL-SE}
\mathrm{SE}_j^{\rm{(ul)}} = K  \zeta^{\rm{(ul)}} \! \left( 1-\frac{B}{S} \right) \log_2 \! \left(1+ \frac{1}{I_{j}^{\mathrm{scheme}} } \right) \, \text{[bit/s/Hz/cell] }
\end{equation}
where the interference term 
\begin{align} \notag
&I_{j}^{\mathrm{scheme}} = \fracSum{l \in \mathcal{L}_j(\beta) \setminus \{ j \}}  \left( \mu^{(2)}_{jl} + \frac{\mu^{(2)}_{jl}-\left( \mu^{(1)}_{jl} \right)^2}{G^{\mathrm{scheme}}} \right)  \\ & \quad +  \frac{ \left( \fracSum{l \in \mathcal{L} }\mu_{jl}^{(1)} Z_{jl}^{\mathrm{scheme}} + \frac{\sigma^2}{  \rho } \right) \left(\fracSum{\ell \in \mathcal{L}_j(\beta) } \mu_{j \ell}^{(1)}  + \frac{\sigma^2}{ B \rho} \right)  }{G^{\mathrm{scheme}}}    \label{eq:achievable-SINR-ULscheme}
\end{align}
depends on the receive combining scheme through $G^{\mathrm{scheme}}$ and $Z_{jl}^{\mathrm{scheme}}$. MR combining is obtained by $G^{\mathrm{MR}}=M$ and $Z_{jl}^{\mathrm{MR}}=K$, while ZF combining is obtained by $G^{\mathrm{ZF}}=M-K$ and 
\begin{equation}
Z_{jl}^{\mathrm{ZF}}= \begin{cases}
K \left( \!
1-
\frac{ \mu_{j l}^{(1)}  }{  \fracSum{\ell \in \mathcal{L}_j(\beta) }  \mu_{j \ell}^{(1)} + \frac{\sigma^2}{ B \rho} }
\! \right)
& \textrm{if } l \in \mathcal{L}_j(\beta), \\
K & \textrm{if } l \not \in \mathcal{L}_j(\beta).
\end{cases}
\end{equation}
The following notation was used:
\begin{align} \label{eq:mu-definition}
\mu^{(\omega)}_{jl} &= \mathbb{E}_{\vect{z}_{lm}} \left\{ \left( \frac{d_j(\vect{z}_{lm}) }{ d_l(\vect{z}_{lm})} \right)^{\omega} \right\} \quad \text{for} \,\,\, \omega=1,2.
\end{align}
\end{theorem}
\begin{IEEEproof}
The proof is given in the appendix.
\end{IEEEproof}

{ The closed-form SE expressions in Theorem \ref{theorem-achievable-rate-conventional} are lower bounds on the ergodic capacity and slightly more conservative than the non-closed-form bound in Lemma \ref{lemma:SE}; see Section \ref{subsec:impact-system-parameters} for a numerical comparison. We stress that the closed-form SEs are only functions of the pilot allocation and the propagation parameters $\mu^{(1)}_{jl}$ and $\mu^{(2)}_{jl}$ defined in \eqref{eq:mu-definition}.} The latter two are the average ratio between the channel variance to BS $j$ and the channel variance to BS $l$, for an arbitrary UE in cell $l$, and the second-order moment of this ratio, respectively. These parameters are equal to 1 for $j=l$ and otherwise go to zero as the distance between BS $j$ and cell $l$ increases. The SE expression manifests the importance of pilot allocation, {since the interference term in \eqref{eq:achievable-SINR-ULscheme} contains summations that only consider the cells that use the same pilots as cell $j$.}

{The first term in \eqref{eq:achievable-SINR-ULscheme} describes the pilot contamination, while the second term is the inter-user interference. The difference between MR and ZF is that the latter scheme cancels some interference through $Z_{jl}^{\mathrm{scheme}}$, at the price of reducing the array gain $G^{\mathrm{scheme}}$ from $M$ to $M-K$.}

ZF combining only actively suppresses intra-cell interference, while the inter-cell interference is passively suppressed just as in MR combining. Further interference rejection can be achieved by coordinating the combining across cells, such that both intra-cell and inter-cell interference are actively suppressed by the receive combining.
We propose a new \emph{full-pilot zero-forcing (P-ZF) combining}, defined as
\begin{equation} \label{eq:def-PZF}
\vect{g}_{jk}^{\mathrm{P}\text{-}\mathrm{ZF}} = \widehat{\vect{H}}_{\mathcal{V},j} \big( \widehat{\vect{H}}_{\mathcal{V},j}^{\Htran} \widehat{\vect{H}}_{\mathcal{V},j} \big)^{-1} \vect{e}_{i_{jk}}.
\end{equation}
In contrast to the conventional ZF in \eqref{eq:def-ZF}, which only orthogonalize the $K$ intra-cell channels in $\widehat{\vect{H}}_{\mathcal{V},j} \vect{E}_j$, P-ZF exploits that all the $B$ estimated channel directions in $\widehat{\vect{H}}_{\mathcal{V},j}$ are known at BS $j$ and orthogonalizes all these directions to also mitigate parts of the inter-cell interference; a similar downlink concept was proposed in \cite{Huh2012a}. The cost is a loss in array gain of $B$, instead of $K$ as with conventional ZF. There is no signaling between BSs in this coordinated multipoint (CoMP) scheme---BS $j$ estimates $\widehat{\vect{H}}_{\mathcal{V},j}$ from the UL pilot signaling---and thus the P-ZF scheme is fully distributed and scalable.
Achievable SEs with P-ZF are given by the following theorem.

\begin{theorem} \label{theorem-achievable-rate-pilotbased}
 
Let $\mathcal{L}_l(\beta) \subset \mathcal{L}$ be the subset of cells that uses the same pilots as cell $l$.
In the UL, an achievable SE in cell $j$ with P-ZF combining is given by \eqref{eq:UL-SE} for $G^{\mathrm{P}\text{-}\mathrm{ZF}}=M-B$ and 
\begin{equation}
Z_{jl}^{\mathrm{P}\text{-}\mathrm{ZF}}= K \left( \!
1-
\frac{  \mu_{j l}^{(1)} }{  \fracSum{\ell \in \mathcal{L}_l(\beta) } \mu_{j \ell}^{(1)} + \frac{\sigma^2}{ B \rho} }
\! \right).
\end{equation}
\end{theorem}
\begin{IEEEproof}
The proof is given in the appendix.
\end{IEEEproof}

{ The SE expressions were derived assuming that $M$ and $K$ are the same in all cells, for notational brevity. However, the results in this section are straightforward to extend to cell-specific $M$ and $K$ values.}

\subsection{Achievable DL Spectral Efficiencies}

The channel estimates from Lemma \ref{lemma:LMMSE-estimation} are also used for linear precoding  in the DL, where the $M$ channel inputs are utilized to make each data signal add up (semi-)coherently at its desired UE and to suppress the interference caused to other UEs. Recall from \eqref{eq:system-model-DL} that $\vect{w}_{jk} \in \mathbb{C}^{M}$ is the precoding vector associated with UE $k$ in cell $j$. We express these precoding vectors as
\begin{equation}
\vect{w}_{jk} = \sqrt{ \frac{ q_{jk}}{ \mathbb{E}_{\{\vect{h}\}} \{\| \check{\vect{g}}_{jk} \|^2 \}}} \check{\vect{g}}_{jk}^*
\end{equation}
where the average transmit power $q_{jk}\geq 0 $ is a function of the UE positions, but not the instantaneous channel realizations. The vector $\check{\vect{g}}_{jk} \in \mathbb{C}^{M}$ defines the spatial directivity of the transmission and is based on the acquired CSI; the normalization with the average squared norm $\mathbb{E}_{\{\vect{h}\}} \{\| \check{\vect{g}}_{jk} \|^2 \}$ gives the analytic tractability that enables the following results.\footnote{Conventionally, the power is normalized by $\| \check{\vect{g}}_{jk} \|^2$ instead of $\mathbb{E}_{\{\vect{h}\}} \{\| \check{\vect{g}}_{jk} \|^2 \}$ in multi-user MIMO systems \cite{Bjornson2013d}, but the difference is small in massive MIMO since  $| \mathbb{E}_{\{\vect{h}\}} \{\| \check{\vect{g}}_{jk}\|^2  \} - \| \check{\vect{g}}_{jk} \|^2 |/M \rightarrow 0$ as $M \rightarrow \infty$, for most precoding schemes.}

\begin{lemma} \label{lemma:SE-DL}
In the DL, an ergodic achievable SE {of an arbitrary} UE $k$ in cell $j$ is
\begin{equation}
\zeta^{\rm{(dl)}} \left( 1-\frac{B}{S} \right) \mathbb{E}_{\{\vect{z}\}} \left\{ \log_2(1+ \mathrm{SINR}_{jk}^{\rm{(dl)}}) \right\} \quad \text{[bit/s/Hz]}
\end{equation}
with the effective SINR, $\mathrm{SINR}_{jk}^{\rm{(dl)}}$, given by
\begin{equation} \label{eq:SINR-value-DL}
\frac{ q_{jk} \frac{| \mathbb{E}_{\{\vect{h}\}}\{ \check{\vect{g}}_{jk}^{\Htran} \vect{h}_{jjk} \} |^2}{ \mathbb{E}_{\{\vect{h}\}} \{\| \check{\vect{g}}_{jk} \|^2 \}} }{ \fracSum{l \in \mathcal{L}} \fracSumtwo{m=1}{K} q_{lm} \frac{ \mathbb{E}_{\{\vect{h}\}} \{ |\check{\vect{g}}_{lm}^{\Htran} \vect{h}_{ljk} |^2  \} }{ \mathbb{E}_{\{\vect{h}\}} \{\| \check{\vect{g}}_{lm} \|^2 \} }- q_{jk} \frac{| \mathbb{E}_{\{\vect{h}\}}\{ \check{\vect{g}}_{jk}^{\Htran} \vect{h}_{jjk} \} |^2}{ \mathbb{E}_{\{\vect{h}\}} \{\| \check{\vect{g}}_{jk} \|^2 \}}  + \sigma^2 }.
\end{equation}
\end{lemma}
\begin{IEEEproof}
This follows from the same procedures as the proof of Lemma \ref{lemma:SE}.
\end{IEEEproof}

Note that Lemma \ref{lemma:SE-DL} takes into account the fact that each UE only knows the expectations in \eqref{eq:SINR-value-DL} and not the instantaneous channels (see \cite[Theorem 1]{Jose2011b} for more details).

The precoding can be designed in a variety of ways. The next theorem shows that there is a strong connection between transmit precoding in the DL and receive combining in the UL.

\begin{theorem} \label{theorem:DLUL-duality}
Let $\{ \vect{g}_{jk}^{\mathrm{scheme}} \}$ be the set of receive combining vectors used in the UL. Then, there exist a DL
power control policy $\{ q_{jk} \}$, with $\fracSum{j \in \mathcal{L}} \fracSumtwo{k=1}{K}  q_{jk} = \fracSum{j \in \mathcal{L}} \fracSumtwo{k=1}{K}  p_{jk}$, for which
\begin{equation}
\mathrm{SINR}_{jk}^{\rm{(dl)}} = \mathrm{SINR}_{jk}^{\rm{(ul)}}
\end{equation}
by using $\check{\vect{g}}_{jk} = \vect{g}_{jk}^{\mathrm{scheme}}$ for all $j$ and $k$. Consequently, an achievable SE in the DL of cell $j$ is
\begin{equation} \label{eq:asymptotic-SE-DL}
\mathrm{SE}_j^{\rm{(dl)}} = K \zeta^{\rm{(dl)}} \! \left( 1-\frac{B}{S} \right) \log_2 \! \left(1+ \frac{1}{I_{j}^{\mathrm{scheme}} } \right) \, \text{[bit/s/Hz/cell]}
\end{equation}
where the interference term $I_{j}^{\mathrm{scheme}}$ is the same as in the UL (for MR, ZF, or P-ZF).
\end{theorem}
\begin{IEEEproof}
The proof is given in the appendix.
\end{IEEEproof}

This theorem shows that the SINRs that are achieved in the UL are also achievable in the DL, by selecting the power control coefficients $\{ q_{jk} \}$ properly. The total transmit power is the same, but is allocated differently over the UEs. This is a consequence of the uplink-downlink duality \cite{Boche2002a}, conventionally considered for single-cell systems with perfect CSI, which is applicable also in our general multi-cell massive MIMO setup with estimated CSI. The exact expression for the power control coefficients is only given in the proof, since the main purpose of Theorem \ref{theorem:DLUL-duality} is the fact that equal UL/DL performance is possible, which allows for joint analysis in what follows.

Motivated by Theorem \ref{theorem:DLUL-duality}, this paper considers three types of linear precoding vectors: MR precoding which amplifies the desired signal by setting $\check{\vect{g}}_{jk} = \vect{g}_{jk}^{\mathrm{MR}}$; ZF precoding that actively rejects intra-cell interference by setting $\check{\vect{g}}_{jk} = \vect{g}_{jk}^{\mathrm{ZF}}$; and P-ZF precoding that actively rejects both intra- and inter-cell interference by setting $\check{\vect{g}}_{jk} = \vect{g}_{jk}^{\mathrm{P}\text{-}\mathrm{ZF}}$. We stress that P-ZF precoding is a fully distributed coordinated beamforming scheme tailored to massive MIMO systems, since each BS only uses locally estimated CSI.

\subsection{Finite and Asymptotic Analysis}

Based on Theorems \ref{theorem-achievable-rate-conventional}--\ref{theorem:DLUL-duality}, the sum of the per-cell achievable SEs in the UL and DL are given by the following corollary.

\begin{corollary} \label{corollary-achievable-rate-DLUL}  
Looking jointly at the UL and DL, an achievable SE in cell $j$ is
\begin{equation} \label{eq:UL-DL-rate}
\begin{split}
&\mathrm{SE}_j = \mathrm{SE}_j^{\rm{(ul)}} + \mathrm{SE}_j^{\rm{(dl)}} \\ &= K \left( 1-\frac{B}{S} \right) \log_2 \left(1+ \frac{1}{I_{j}^{\mathrm{scheme}} } \right) \, \text{[bit/s/Hz/cell]}
\end{split}
\end{equation}
where the interference term $I_{j}^{\mathrm{scheme}}$ for UE $k$ is given 
Theorem \ref{theorem-achievable-rate-conventional} for MR and ZF and in Theorem \ref{theorem-achievable-rate-pilotbased} for P-ZF. This SE can be divided between the UL and DL arbitrarily using any positive fractions $\zeta^{\rm{(ul)}}$ and $\zeta^{\rm{(dl)}}$, with $\zeta^{\rm{(ul)}} + \zeta^{\rm{(dl)}} = 1$.
\end{corollary}

This is a convenient result that allows us to analyze and optimize the SE of the network as a whole, without having to separate the UL and DL. {Since it is hard to gain further insights from the structure of the SE expression in \eqref{eq:UL-DL-rate}, we analyze it for a particular network topology in Section \ref{sec:optimization-hexagonal}. In the remainder of this section, we consider the limit of a large number of antennas.}

\begin{corollary} 
Let $\mathcal{L}_j(\beta) \subset \mathcal{L}$ be the subset of cells that uses the same pilots as cell $j$.
When $M \rightarrow \infty$ (with $K,B \leq S < \infty$), the effective SINRs with MR, ZF, and P-ZF converge to the same limit:
\begin{equation} \label{eq:asymptotic-SINR}
\frac{1}{I_{j}^{\mathrm{MR}} }, \frac{1}{I_{j}^{\mathrm{ZF}} }, \frac{1}{I_{j}^{\mathrm{P}\text{-}\mathrm{ZF}} } \rightarrow \frac{ 1 }{ \fracSum{l \in \mathcal{L}_j(\beta) \setminus \{j \} }  \mu^{(2)}_{jl}    }.
\end{equation}
\end{corollary}

{ The ultimate effect of pilot contamination is very clear in \eqref{eq:asymptotic-SINR}, since only the cells that interfered with cell $j$ during pilot transmission (i.e., cells with indices in the set $\mathcal{L}_j(\beta) \setminus \{j \} $) affect the asymptotic limit. To maximize the asymptotic SINR in \eqref{eq:asymptotic-SINR}, one should place the cells with large $\mu^{(2)}_{jl}$ in different subsets (i.e., $\mathcal{L}_j(\beta) \cap \mathcal{L}_l(\beta) = \emptyset$) so that these cells use different pilots. The asymptotic limit can  be used as follows to find the optimal $K$.}

\begin{corollary} \label{cor:SE-maximization} {
Let $\mathcal{L}_j(\beta) \subset \mathcal{L}$ be the subset of cells that uses the same pilots as cell $j$.}
The SE in cell $j$ approaches
\begin{equation} \label{eq:asymptotic-SE}
\mathrm{SE}_j^{\infty} = K \left( 1-\frac{K \beta}{S} \right) \log_2 \bigg(1+ \frac{ 1 }{ \sum_{l \in \mathcal{L}_j(\beta) \setminus \{ j\} } \mu^{(2)}_{jl} } \bigg)
\end{equation}
when $M \rightarrow \infty$. This SE is maximized jointly for all cells when the number of scheduled UEs is either $K^\star = \left\lfloor \frac{S}{2 \beta} \right\rfloor$ or $K^\star = \left\lceil \frac{S}{2 \beta} \right\rceil$ (i.e., one of the closest integers to $\frac{S}{2 \beta} $).
\end{corollary}
\begin{IEEEproof}
The logarithmic part of \eqref{eq:asymptotic-SE} is independent of $K$, while the concave pre-log factor $K \left( 1-\frac{K \beta}{S} \right)$ is maximized by $K = \frac{S}{2 \beta}$. The concavity implies that the optimal integer $K^*$ is one of the closest integers to $\frac{S}{2 \beta}$.
\end{IEEEproof}

Corollary \ref{cor:SE-maximization} is a main contribution of this paper and proves that the number of scheduled UEs should be proportional to the frame length $S$ (when $M$ is large enough); for example, we get $K^\star=\frac{S}{2}$ for $\beta = 1$ and $K^\star=\frac{S}{6}$ for $\beta = 3$. Since both $S=200$ and $S = 10000$ are reasonable coherence block lengths in practice, depending on the UE mobility and propagation environment, this means that we should schedule between tens and several thousands of UEs for simultaneous transmission in order to be optimal. This is only possible if the UE selection policy is scalable and there is a high load of UEs.  If $K^\star = \frac{S}{2 \beta}$ is an integer, the asymptotically optimal SE is
\begin{equation} \label{eq:asymptotic-SE-optimized}
\mathrm{SE}_j^{\infty} = \frac{S}{4 \beta} \log_2 \bigg(1+ \frac{ 1 }{ \sum_{l \in \mathcal{L}_j(\beta) \setminus \{ j\} } \mu^{(2)}_{jl} } \bigg)
\end{equation}
and increases linearly with the frame length $S$ (in the large-$M$ regime).

Interestingly, the asymptotically optimal scheduling gives $B = \frac{S}{2}$ for any $\beta$, which means that \emph{half} the frame is allocated to pilot transmission. This extraordinary fact was initially conjectured in \cite{Marzetta2010a} for $\beta =1$. The rationale is that the SE gain from adding an extra UE outweighs the pre-log loss at the existing UEs if at least half the frame is used for data (a criterion independent of $\beta$). The asymptotically optimal $\beta$ cannot be computed in closed-form, but we notice that a larger $\beta$ leads to fewer interferers in $\mathcal{L}_j(\beta) $ and also reduces the pre-log factor; hence, a larger $\beta$ brings SINR improvements until a certain point where the pre-log loss starts to dominate.

At first sight, these results bear some similarity  with the results in \cite{Hassibi2003a} and \cite{Zheng2002a} for block-fading noncoherent point-to-point (P2P) MIMO channels, where the maximal degrees of freedom (DoF) are $\frac{S}{4}$ and are achieved by having $\frac{S}{2}$ transmit/receive antennas and using pilot signals of the same length. The fundamental difference is that the DoF concept for P2P MIMO channels, where unbounded SE is achieved at high SNRs, does not apply to cellular networks \cite{Lozano2013a}. Instead, the pre-log factor $\frac{S}{4 \beta}$ in \eqref{eq:asymptotic-SE-optimized} may be interpreted as the relative improvement in SE that can be achieved by aggressive scheduling of UEs in massive MIMO systems.

We have now established the asymptotically optimal number of scheduled UEs, as $M \rightarrow \infty$.
Next, we investigate the impact on practical systems with finite $M$ for a certain network topology.

\section{Optimizing Number of UEs in Hexagonal Networks}
\label{sec:optimization-hexagonal}

The concept of cellular communications has been around for decades \cite{Macdonald1979a}. Although practical deployments have irregular cells, it is common practice to establish general properties by analyzing symmetric networks where the cells are regular polygons \cite{Cox1982a}; in particular, hexagons.

\begin{figure}[!t]
\begin{center} 
\includegraphics[width=.8\columnwidth]{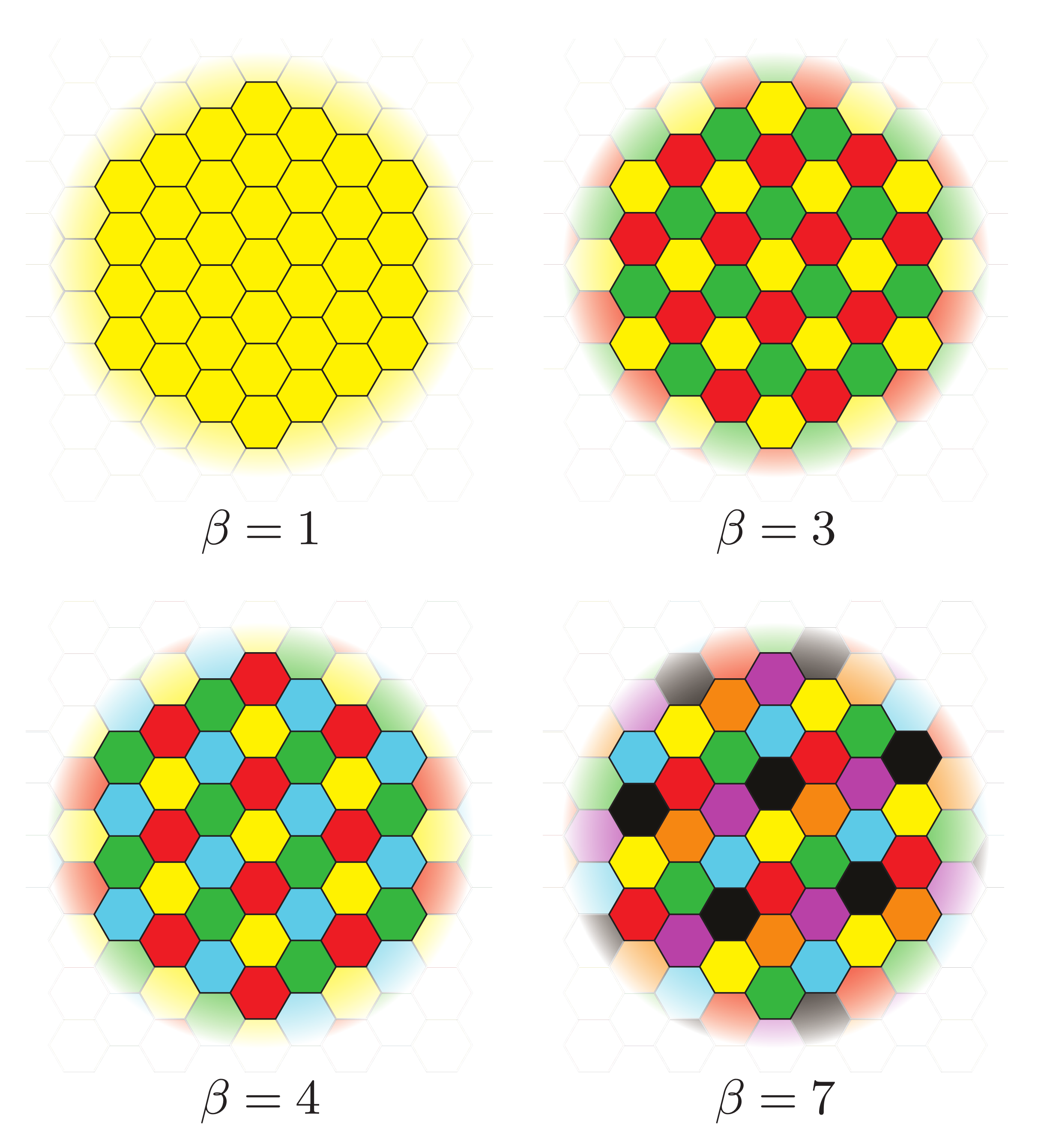}
\end{center}\vskip-3mm
\caption{Part of a hexagonal network, colored for different pilot reuse factors $\beta$.} \label{figure_hexagonal} \vskip-4mm
\end{figure}

\begin{figure}[!t]
\begin{center} 
\includegraphics[width=.5\columnwidth]{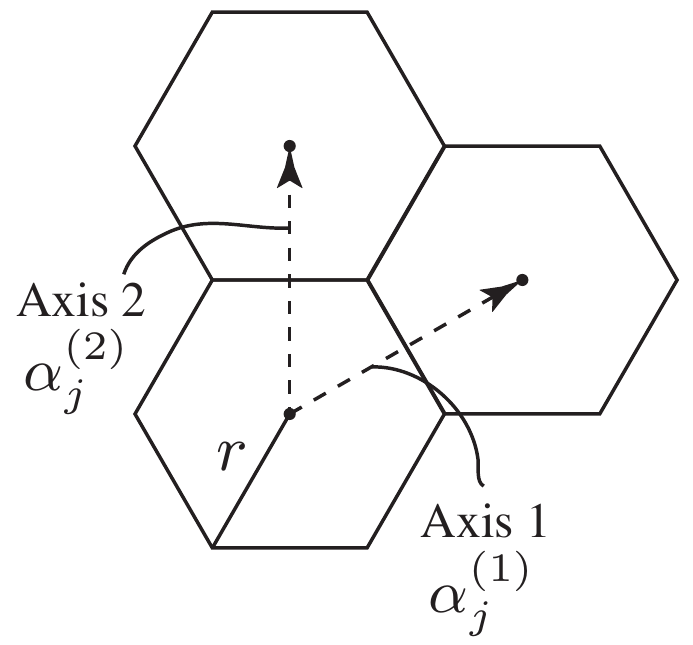}
\end{center}\vskip-3mm
\caption{The coordinate system for a hexagonal grid.} \label{figure_hexagonal_coordinates} \vskip-6mm
\end{figure}

In this section, we consider the symmetric network topology depicted in Fig.~\ref{figure_hexagonal} with hexagonal cells. All the time/frequency resources allocated to payload data transmission are used in all the cells. However, inspired by \cite{Huh2012a}, we consider pilot books of size $B = \beta K$ to allow for non-universal pilot reuse that mitigates the pilot contamination from neighboring cells.

The hexagonal grid is infinitely large, to avoid edge effects and to give all cells the same properties. The cell radius is denoted by $r>0$ and is the distance from the cell center to the corners. Each cell can be uniquely indexed by a pair of integers $\alpha_j^{(1)},\alpha_j^{(2)} \in \mathbb{Z}$, where $\mathbb{Z}$ is the set of integers. This integer pair specifies the location of BS $j$ \cite{Macdonald1979a}:
\begin{equation}
\vect{b}_j = \sqrt{3}
\begin{bmatrix}
\sqrt{3}r/2 \\
r/2
\end{bmatrix} \alpha_j^{(1)}
+\begin{bmatrix}
0 \\
\sqrt{3}r
\end{bmatrix} \alpha_j^{(2)}  \in \mathbb{R}^2.
\end{equation}
The coordinate system imposed by $\alpha_j^{(1)}$ and $\alpha_j^{(2)}$ is illustrated in Fig.~\ref{figure_hexagonal_coordinates}. Every cell on the hexagonal grid has 6 interfering cells in the first surrounding tier, 12 in the second tier, etc. As shown in the early works on hexagonal networks \cite{Macdonald1979a,Cox1982a}, this limits which pilot reuse factors that give symmetric reuse patterns: $\beta \in \{1,3,4,7,9,12,13, \ldots \}$.

Our simulations consider a classic pathloss model where the variance of the channel attenuation in \eqref{eq:channel-distribution} is $d_j(\vect{z}) = \frac{C}{\| \vect{z} - \vect{b}_j\|^{\kappa}}$, where $\|\cdot\|$ is the Euclidean norm, $C>0$ is a reference value, and $\kappa\geq 2$ is the pathloss exponent. These assumptions allow us to compute $\mu^{(\omega)}_{jl}$ in \eqref{eq:mu-definition} as
\begin{equation} \label{eq:mu-computation}
\mu^{(\omega)}_{jl}  = \mathbb{E}_{\vect{z}_{lm}} \! \left\{ \left( \frac{d_j(\vect{z}_{lm}) }{ d_l(\vect{z}_{lm})} \right)^{\omega} \right\} = \mathbb{E}_{\vect{z}_{lm}} \! \left\{ \left(\frac{\| \vect{z}_{lm} \!-\! \vect{b}_l\|}{\| \vect{z}_{lm} \!-\! \vect{b}_j\|} \right)^{\kappa \omega} \right\}
\end{equation}
for any UE distributions in the cells. {We note that $C$ and $r$ cancel out in \eqref{eq:mu-computation}, if the UE distributions in each cell are independent of $C$ and $r$.} Since the power control makes the SEs in Theorems \ref{theorem-achievable-rate-conventional}--\ref{theorem:DLUL-duality} independent of the UEs' positions, we only need to define the parameter ratio $\rho/\sigma^2$; that is, the average SNR (over fading) between any UE and any antenna at its serving BS.

\begin{figure}[th!]
\begin{subfigure}[b]{\columnwidth}
\includegraphics[width=\columnwidth]{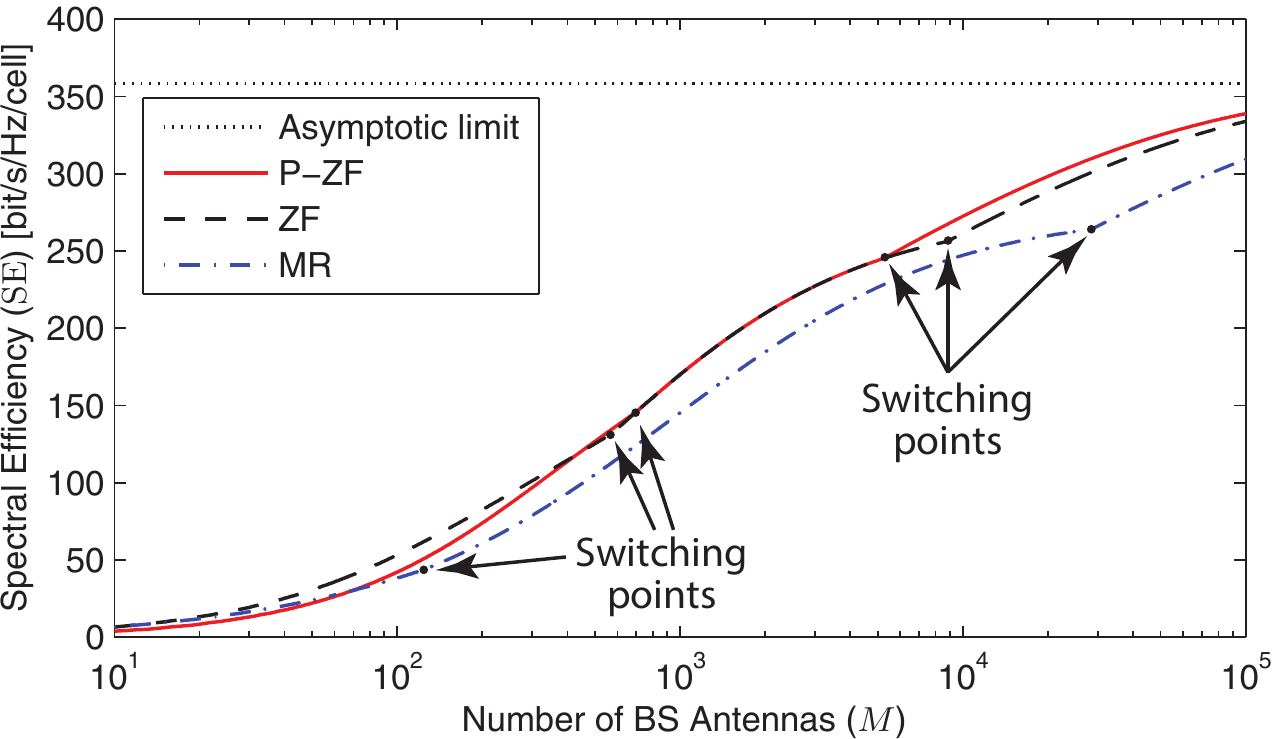} \vspace{-4mm}
\caption{Optimized SE per cell.} \label{figure_SE_mean} \vspace{4mm}
\end{subfigure}
\begin{subfigure}[b]{\columnwidth}
\includegraphics[width=\columnwidth]{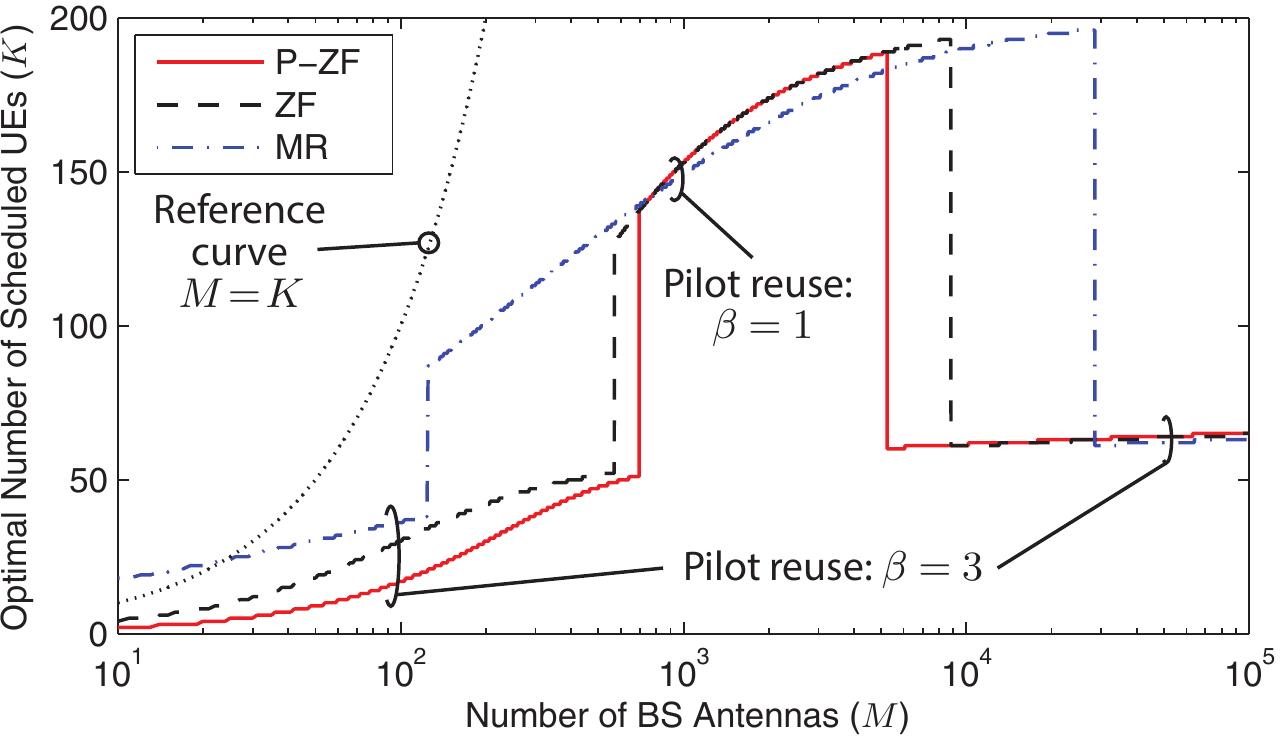} \vspace{-4mm}
\caption{Corresponding optimal number of UEs: $K^\star$.} \label{figure_K_mean}
\end{subfigure} \vspace{-3mm}
\caption{Simulation of optimized SE, as a function of $M$, with average inter-cell interference.} \label{figure_simulation_mean}
 \end{figure}

\begin{figure}[th!]
\begin{subfigure}[b]{\columnwidth}
\includegraphics[width=\columnwidth]{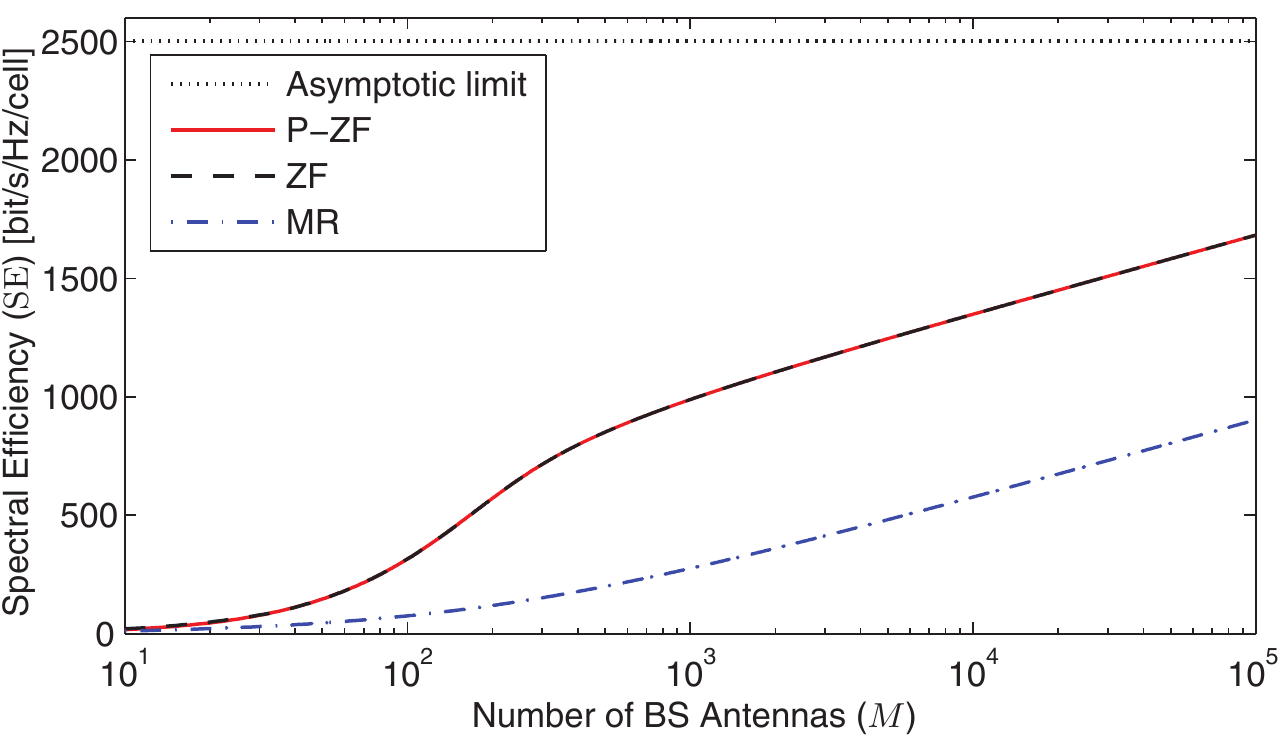} \vspace{-4mm}
\caption{Optimized SE per cell.} \label{figure_SE_best} \vspace{4mm}
\end{subfigure}
\begin{subfigure}[b]{\columnwidth}
\includegraphics[width=\columnwidth]{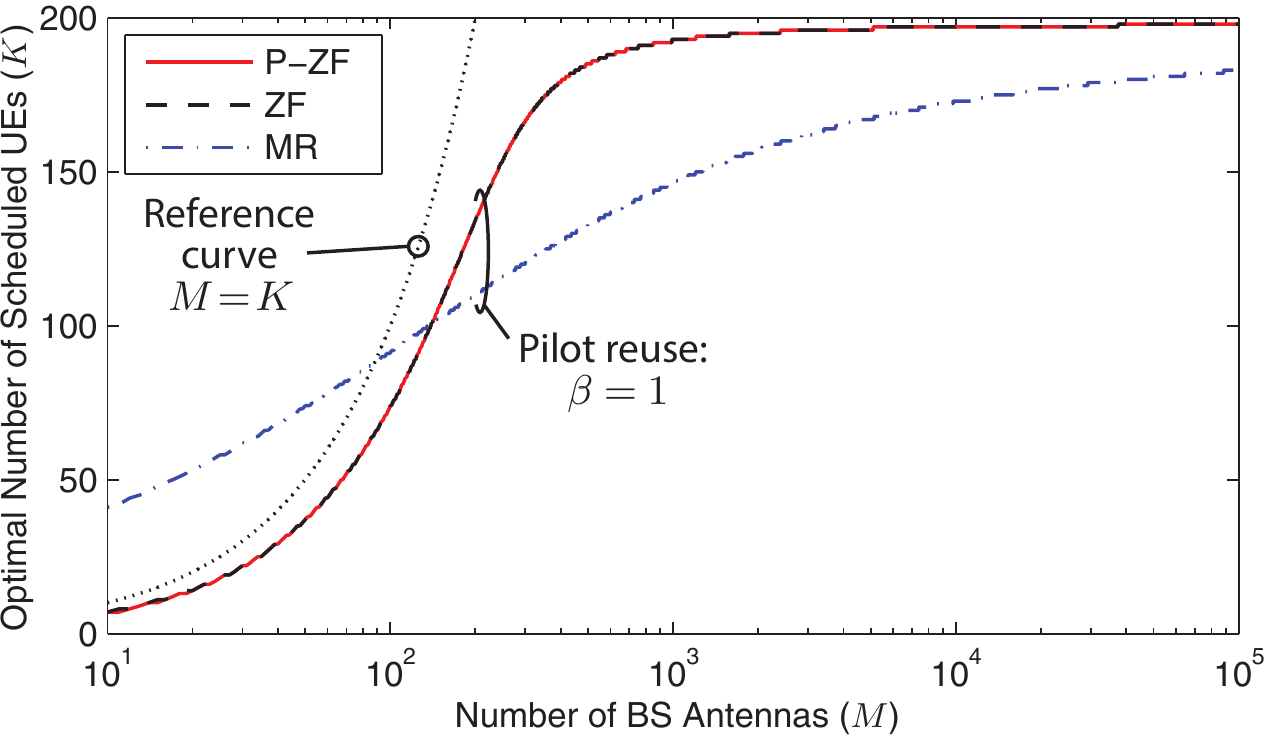} \vspace{-4mm}
\caption{Corresponding optimal number of UEs: $K^\star$.} \label{figure_K_best}
\end{subfigure} \vspace{-3mm}
\caption{Simulation of optimized SE, as a function of $M$, with best-case inter-cell interference.} \label{figure_simulation_best}
 \end{figure}

\begin{figure}[th!]
\begin{subfigure}[b]{\columnwidth}
\includegraphics[width=\columnwidth]{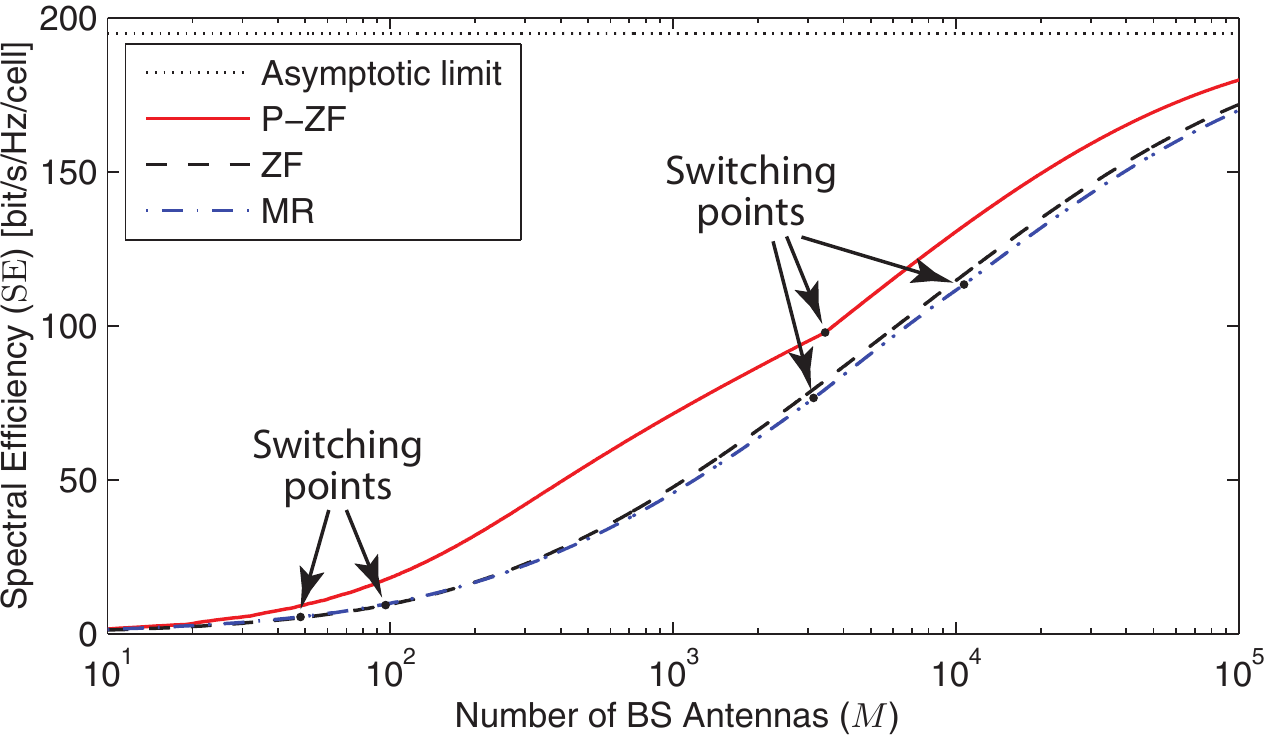} \vspace{-4mm}
\caption{Optimized SE per cell.} \label{figure_SE_worst} \vspace{4mm}
\end{subfigure}
\begin{subfigure}[b]{\columnwidth}
\includegraphics[width=\columnwidth]{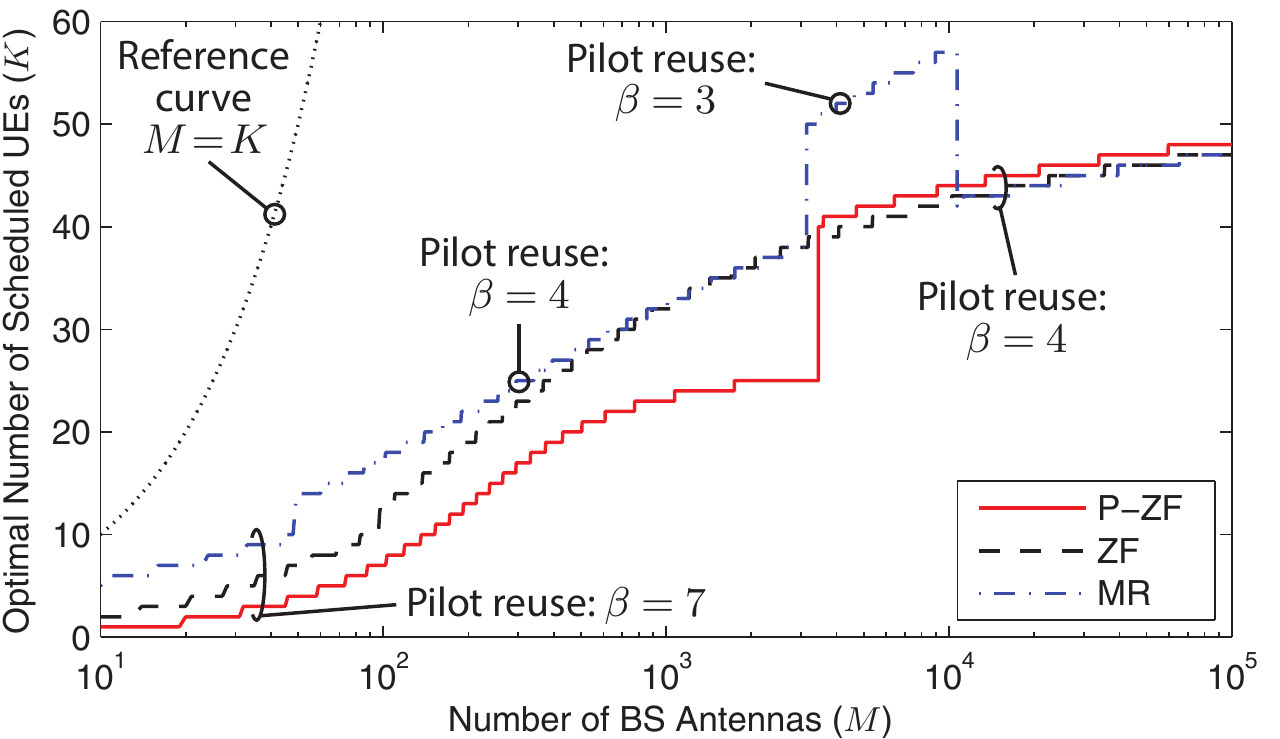} \vspace{-4mm}
\caption{Corresponding optimal number of UEs: $K^\star$.} \label{figure_K_worst}
\end{subfigure} \vspace{-3mm}
\caption{Simulation of optimized SE, as a function of $M$, with worst-case inter-cell interference.} \label{figure_simulation_worst}
 \end{figure}

\subsection{Optimizing SE for Different Interference Levels}

We simulate the SE in an arbitrary cell on the hexagonal grid in Fig.~\ref{figure_hexagonal} and take all non-negligible interference into account. The UEs can be anywhere in the cells, but at least $0.14r$ from the serving BS (this makes the analysis independent of $r$). Since the SE expressions in Section \ref{sec:average-SE} are the same for the UL and DL, except for the fractions $\zeta^{\rm{(ul)}}$ and $\zeta^{\rm{(dl)}}$, we simulate the sum of these SEs and note that it can be divided arbitrarily between the UL and DL. The same linear processing schemes are used in both directions. The simulations consider MR, ZF, and P-ZF precoding/combining, and all results are obtained by computing the closed-form expressions from Section \ref{sec:average-SE} for different parameter combinations. The simulations were performed using Matlab and the code is available for download at \url{https://github.com/emilbjornson/maximal-SE}, which enables reproducibility as well as simple testing of other parameter values.

For each number of antennas, $M$, we optimize the SE with respect to the number of UEs $K$ and the pilot reuse factor $\beta$ (which determine $B = \beta K$) by searching the range of all reasonable integer values. We set the coherence block length to $S = 400$ (e.g., 2 ms coherence time and 200 kHz coherence bandwidth), set the SNR to $\rho/\sigma^2 = 5$ dB, and pick $\kappa = 3.7$ as pathloss exponent.\footnote{A higher pathloss exponent reduces the inter-cell interference, but requires more signal power to maintain a certain SNR.} The impact of changing the different system parameters is considered in Section~\ref{subsec:impact-system-parameters}.

We consider three  propagation environments with different severity of inter-cell interference:
\begin{enumerate}
\item Average case: Averaging over uniform UE locations in all cells.

\item Best case: All UEs in other cells are at the cell edge furthest from BS $j$ (for each $j$).

\item Worst case: All UEs in other cells are at the cell edge closest to BS $j$ (for each $j$).

\end{enumerate}
The corresponding values of the parameters $\mu^{(1)}_{jl}$ and $\mu^{(2)}_{jl}$ were computed by Monte-Carlo simulations with $10^6$ UE locations in each cell.

The best case is overly optimistic since the desirable UE positions in the interfering cells are different with respect to different cells. However, it gives an upper bound on what is achievable by coordinated scheduling across cells. The worst case is overly pessimistic since the UEs cannot all be at the worst locations, with respect to all other cells, at the same time. The average case is probably the most applicable in practice, where the averaging comes from UE mobility, scheduling, and random switching of pilot sequences between the UEs in each cell. Results for the average case are shown in Fig.~\ref{figure_simulation_mean}, the best case in Fig.~\ref{figure_simulation_best}, and the worst case in Fig.~\ref{figure_simulation_worst}. The optimized SE and the corresponding $K^\star$ are shown in (a) and (b), respectively.

The achievable SEs (per cell) are very different between the best case interference and the two other cases---this confirms the fact that results from single-cell analysis of massive MIMO is often not applicable to multi-cell cases (and vice versa). ZF brings much higher SEs than MR under the best case inter-cell interference, since then the potential gain from mitigating intra-cell interference is very high. P-ZF is equivalent to ZF in the best case, but excels under worst case inter-cell interference since it can actively suppress also inter-cell interference. In the realistic average case, the optimized SEs are rather similar for MR, ZF, and P-ZF; particularly in the practical range of $10 \leq M \leq 200$ antennas. In all cases, the largest differences appear when the number of antennas is very large (notice the logarithmic $M$-scales). {At least $M=10^5$ is needed to come close to the asymptotic limit in \eqref{eq:asymptotic-SE-optimized}, which was proved by Corollary \ref{cor:SE-maximization}, and many more antennas are required under best case interference.} Clearly, the asymptotic limits should not be used as performance indicators since unrealistically many antennas are needed for convergence.

As seen from Figs.~\ref{figure_simulation_mean}--\ref{figure_simulation_worst}, the main difference between MR, ZF, and P-ZF is not the values of the optimized SE but how they are achieved; that is, which number of UEs $K^\star$ and which pilot reuse factor $\beta$ that are used. The general behavior is that larger $M$ implies a higher $K^\star$ and a smaller $\beta$, because the channels become more orthogonal with $M$. Since the reuse factor is an integer, $K^\star$ changes non-continuously when $\beta$ is changed; smaller $\beta$ allows for larger $K^\star$, and vice versa. MR schedules the largest number of UEs and switches to a smaller reuse factor at fewer antennas than the other schemes. In contrast, P-ZF schedules the smallest number of UEs and has the highest preference of large reuse factors, since this it can suppress more inter-cell interference in these cases. Simply speaking, MR gives low per-user SEs to many UEs (sometimes more than $M$), while ZF and P-ZF give higher per-user SEs to fewer UEs.

Recall from Corollary \ref{cor:SE-maximization} that $K = \frac{S}{2 \beta}$ becomes the optimal number of UEs as $M \rightarrow \infty$. This property is confirmed by Figs.~\ref{figure_simulation_mean}--\ref{figure_simulation_worst}, since $K^\star \rightarrow 67$ in the average case (where $\beta = 3$), $K^\star \rightarrow 200$ in the best case (where $\beta = 1$), and $K^\star \rightarrow 50$ in the worst case (where $\beta = 4$).

\subsection{Impact of System Parameters}

\label{subsec:impact-system-parameters}

We now focus on the average case of inter-cell interference, due to its practical relevance, and investigate how each system parameter affects the simulation results. We focus on the range $10 \leq M \leq 1000$ antennas, and when other system parameters than $M$ are varied we only consider $M=100$ (medium massive MIMO setup) and $M=500$ (large massive MIMO setup).

\begin{figure}[!t]
\begin{center} 
\includegraphics[width=\columnwidth]{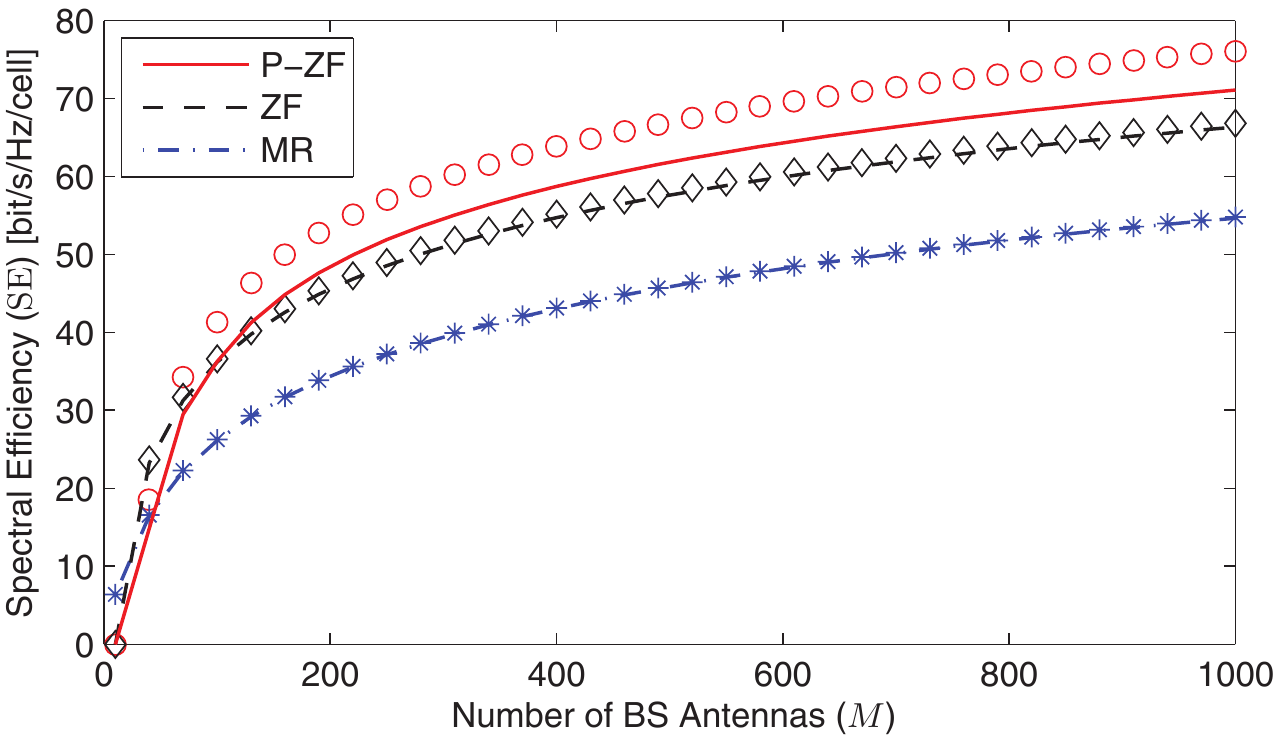}
\end{center}\vskip-3mm
\caption{Per-cell SE for $K=10$. The lines are based on Theorems \ref{theorem-achievable-rate-conventional} and \ref{theorem-achievable-rate-pilotbased} while the markers are computed numerically from Lemma \ref{lemma:SE}.} \label{figure_journal_K10} \vskip-2mm
\end{figure}

We begin by verifying the accuracy of the closed-form expressions in Theorems \ref{theorem-achievable-rate-conventional} and \ref{theorem-achievable-rate-pilotbased}, by comparing the formulas to Monte-Carlo simulations based on Lemma \ref{lemma:SE}. The formulas are exact in the best and worst interference cases, but Fig.~\ref{figure_journal_K10} shows that the interference variations in the average case result in some loss in SE. The figure considers $K=10$ UEs and Monte-Carlo simulations are represented by markers. The MR and ZF formulas in Theorem \ref{theorem-achievable-rate-conventional} are very tight. However, there is a few percent of deviation for P-ZF in Theorem \ref{theorem-achievable-rate-pilotbased}, since a lower bound on the ability of cancel inter-cell interference is used to get a tractable formula. Hence, P-ZF will actually perform slightly better than reported in the simulations in this paper.

\begin{figure}[!t]
\begin{center} 
\includegraphics[width=\columnwidth]{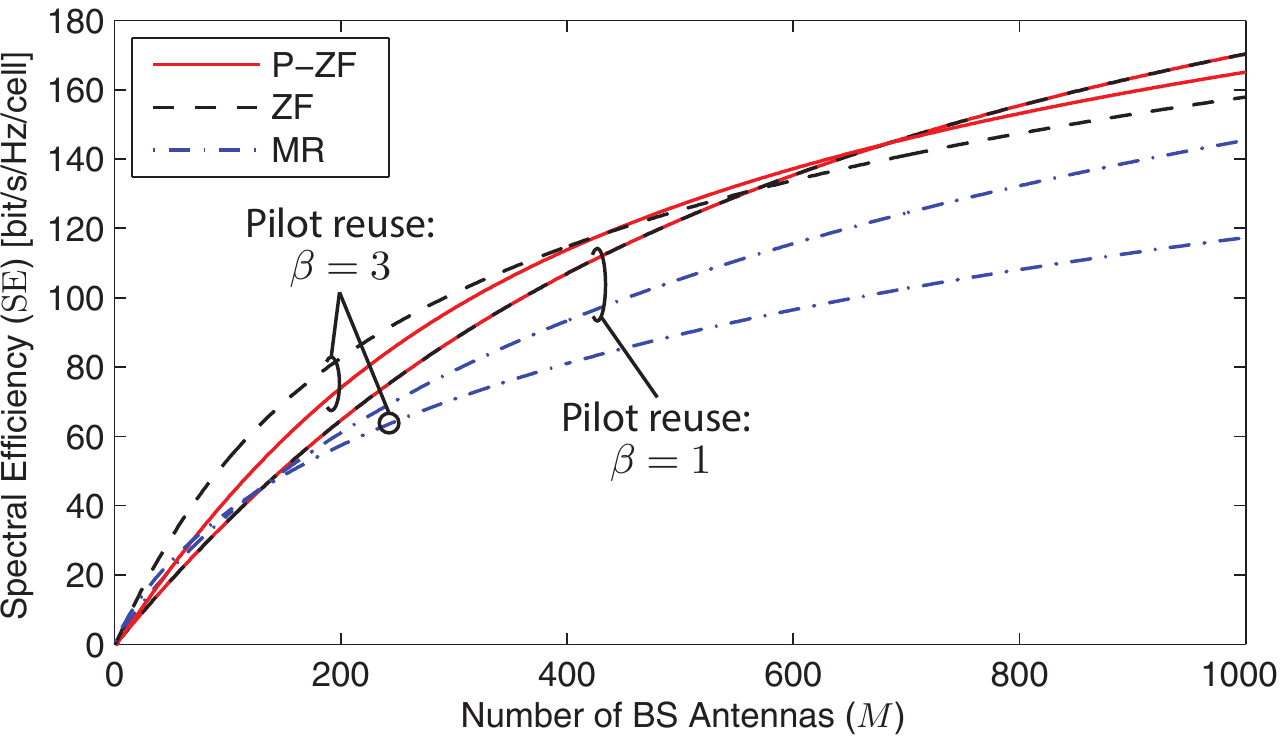}
\end{center}\vskip-3mm
\caption{Impact of changing the pilot reuse factor $\beta$, for a system optimized for high per-cell SE.} \label{figure_journal_reusefactors}\vskip-2mm
\end{figure}

Next, we study the impact of the pilot reuse factor $\beta$ using the formulas from Theorems \ref{theorem-achievable-rate-conventional} and \ref{theorem-achievable-rate-pilotbased}. Fig.~\ref{figure_journal_reusefactors} shows the per-cell SE for $\beta = 1$ and $\beta = 3$, which  provide the highest SEs for $M \leq 1000$. The curves are smooth and there are wide regions around the $\beta$-switching points where both values provide almost equal SEs. This robustness simplifies cell planning and scheduling based on user load.

\begin{figure}[!t]
\begin{center} 
\includegraphics[width=\columnwidth]{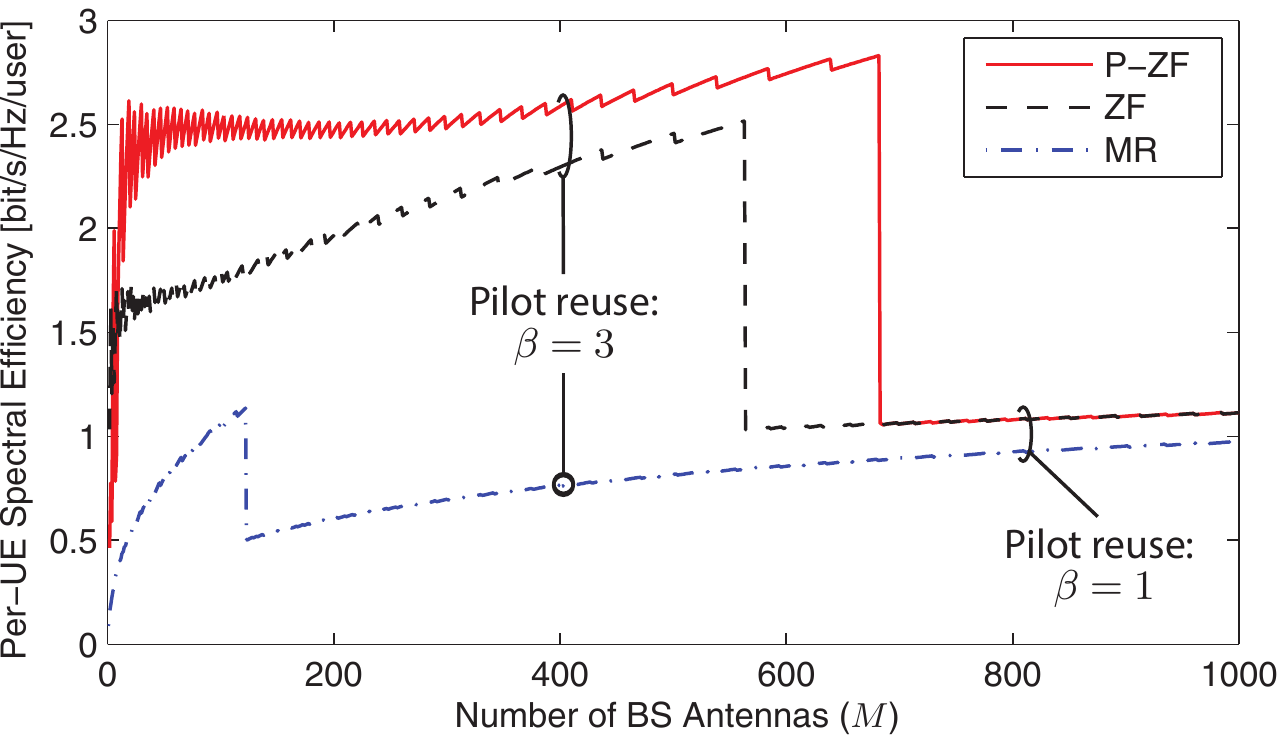}
\end{center}\vskip-3mm
\caption{Achievable SE per UE, for a system optimized for high per-cell SE.} \label{figure_journal_peruser} \vskip-2mm
\end{figure}

Changes in the pilot reuse factor have major impact on the optimal number of UEs and their achievable performance. The SE per UE is shown in Fig.~\ref{figure_journal_peruser} for the operating points that maximize the SE in the cell; this is basically the ratio $\mathrm{SE}/K^\star$, where $\mathrm{SE}$ was given in Fig.~\ref{figure_simulation_mean}(a) and $K^\star$ was given in Fig.~\ref{figure_simulation_mean}(b). We notice that MR gives the lowest SE per scheduled UE, while P-ZF gives the highest SE per scheduled UE. The numbers are around 1 bit/s/Hz for MR, in the range 1--2.5 bit/s/Hz for ZF, and in the range 1--3 bit/s/Hz for P-ZF. Since the pilot signaling consumes between 2 and 40 percent of the frame in this simulation, the payload data need to be encoded with up to 4.5 bit/symbol, which can be achieved by conventional 64-QAM with a 3/4 coding rate. Hence, all the per-user SEs in Fig.~\ref{figure_journal_peruser} are straightforward to implement in practice.

\begin{figure}[!t]
\begin{center} 
\includegraphics[width=\columnwidth]{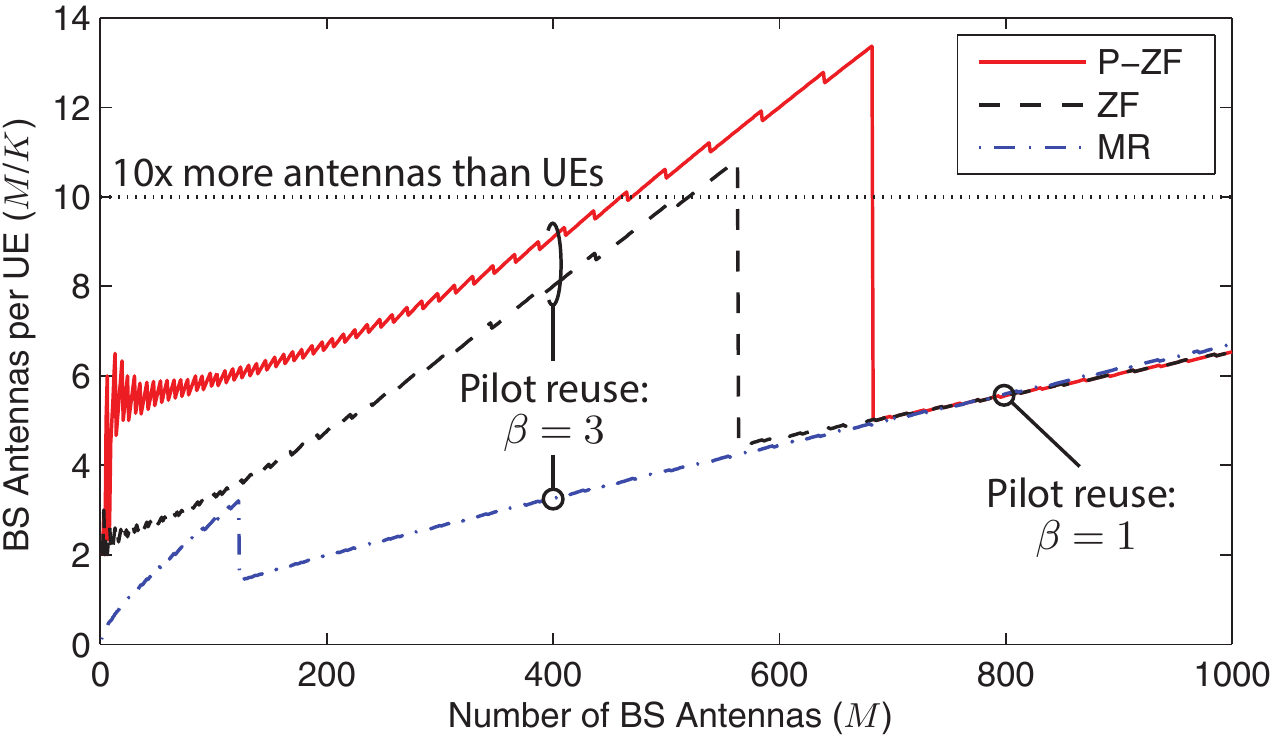}
\end{center}\vskip-3mm
\caption{Number of BS antennas per UE with different processing schemes, for a system optimized for high per-cell SE.} \label{figure_journal_peruser_antennas} \vskip-2mm
\end{figure}

Fig.~\ref{figure_journal_peruser_antennas} shows the ratio $M/K^\star$ for the same scenario as in the previous figures. This ratio can be interpreted as the number of BS antennas per UE \cite{Hoydis2013a}.
There is a common rule of thumb which says that massive MIMO systems should have an order of magnitude more BS antennas than UEs. The operating points that satisfy this guideline are above the horizontal dotted line. This simulation indicates that an optimized system might not follow this guideline; in fact, there is a few occasions where MR even prefers to have $M/K^\star < 1$. Generally speaking, it seems that having 2--8 times more BS antennas than UEs is the range to aim at for practical deployments.

\begin{figure}[!t]
\begin{center} 
\includegraphics[width=\columnwidth]{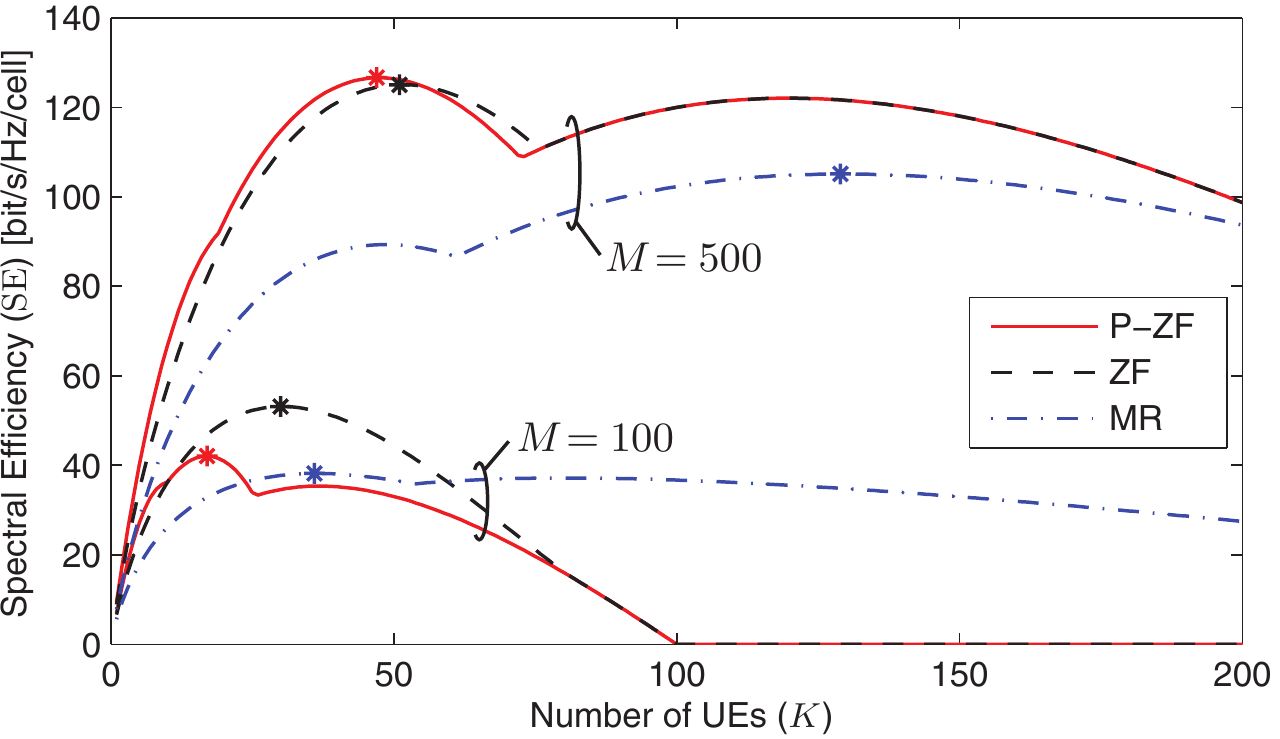}
\end{center}\vskip-3mm
\caption{Achievable per-cell SE as a function of the number of scheduled UEs.} \label{figure_journal_user_SEcell}\vskip-2mm
\end{figure}

Since the cells might not be fully loaded at every time instant, Fig.~\ref{figure_journal_user_SEcell} shows the per-cell SE as a function of the number of scheduled UEs. As noted before, the peak numbers (which are star marked) are at different $K$ for each scheme. If MR, ZF, and P-ZF are compared for a given $K$, the differences between the schemes can either be larger or smaller than at the peak numbers. Although ZF and P-ZF often provide better SE than MR, it is interesting to note that MR is competitive when $K$ is large---both in terms of SE and since its computational complexity scales as $\mathcal{O}(MK)$, while the complexity of ZF and P-ZF scales as $\mathcal{O}(MK^2)$  \cite{Bjornson2015a}.

\begin{figure}[!t]
\begin{center} 
\includegraphics[width=\columnwidth]{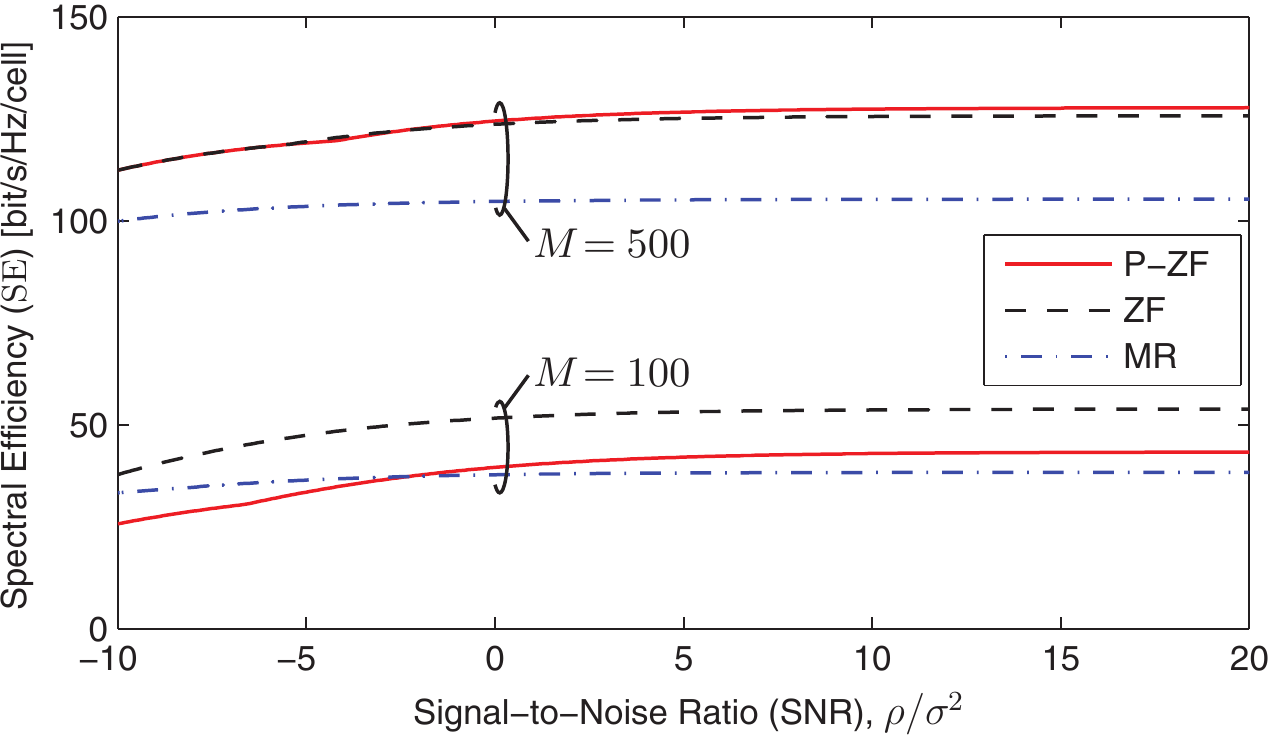}
\end{center}\vskip-3mm
\caption{Impact of SNR variations on the SE.} \label{figure_journal_SNR} \vskip-2mm
\end{figure}

Next, Fig.~\ref{figure_journal_SNR} investigates how the average SNR $\rho/\sigma^2$ affects the results. The SE saturates already at an SNR of 5 dB due to the array gain from coherent processing---this is why 5 dB was used in the previous figures. Massive MIMO can operate also at lower SNRs, but with a performance loss. ZF and P-ZF are particularly sensitive to the SNR level, since the active interference suppression requires a higher CSI estimation quality than simple MR processing.

\begin{figure}[!t]
\begin{center} 
\includegraphics[width=\columnwidth]{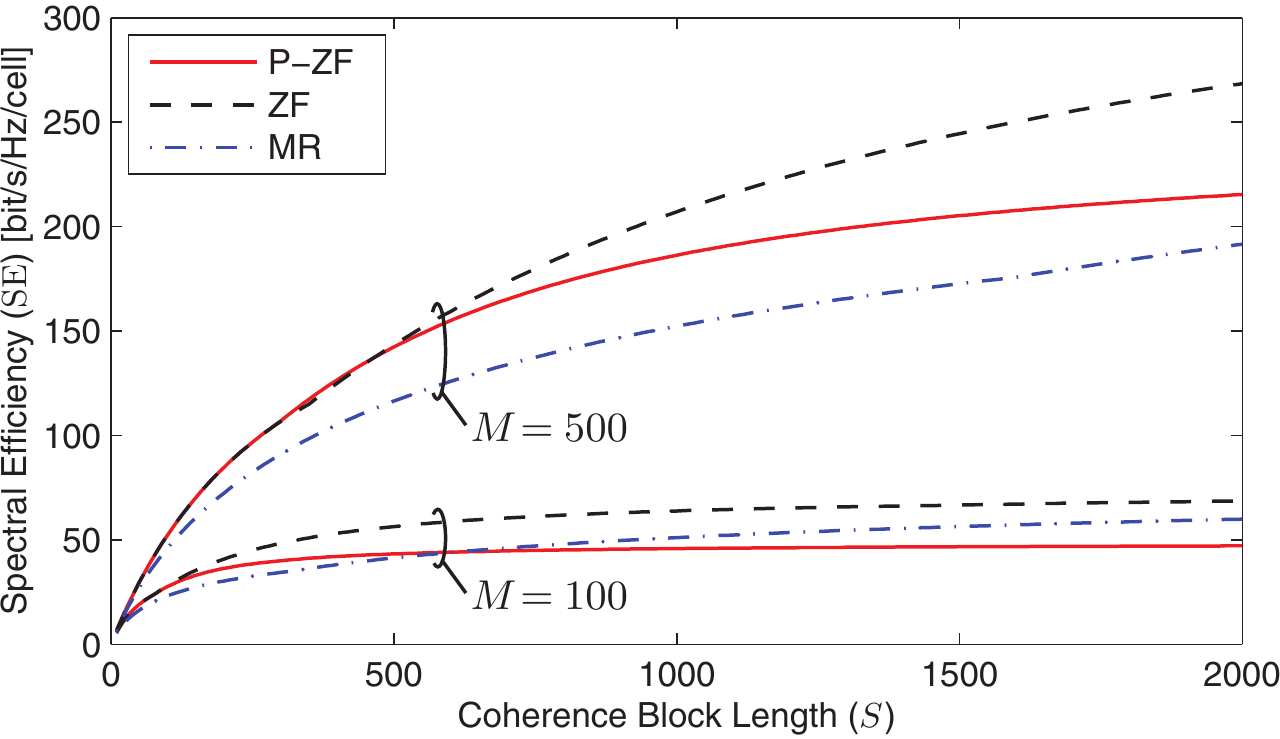}
\end{center}\vskip-3mm
\caption{Per-cell SE as a function of the coherence block length $S$.} \label{figure_journal_coherenceblock} \vskip-2mm
\end{figure}

Finally, Fig.~\ref{figure_journal_coherenceblock} investigates how the length of the coherence block, $S$, affects the per-cell SE. In the case of $M=100$ antennas, the gain of increasing $S$ above 500 is relatively small---the system cannot schedule more UEs since the ratio $M/K$ would then be too small, so the gain mainly comes from reducing the prelog factor ($1-\frac{B}{S}$). However, in the case of $M=500$, the system can utilize an increasing $S$ to schedule more UEs and achieve major improvements in SE. As the number of UEs increases, the part of the intra-cell interference that cannot be rejected due to imperfect CSI becomes the main limiting factor. The benefit of P-ZF then diminishes.

\section{Spectral Efficiencies with Hardware Impairments}
\label{sec:hardware-impairments}

The analytic and numeric analysis in the previous sections have focused on cellular networks where the BSs and UEs are equipped with ideal transceiver hardware, which can radiate any waveform without distortions and which can receive any waveform with infinite resolution. However, practical transceivers inevitably operate with non-linearities in amplifiers, clock drifts in local oscillators, finite-precision ADCs, I/Q imbalance in mixers, and non-ideal analog filters \cite{Schenk2008a,wenk2010mimo,Zhang2012a,Gustavsson2014a}. In this section, we provide a prediction of how these hardware impairments affect the achievable SEs in multi-cell massive MIMO systems. 
We notice that it was recently shown in \cite{Bjornson2014a}, using impairments models developed and evaluated in \cite{Schenk2008a,wenk2010mimo,Zhang2012a}, that the hardware impairments caused by the BS array are negligible in massive MIMO systems, since the desired signals are amplified by the array gain from coherent processing while the distortions add non-coherently. Hence, the hardware impairments in the UE hardware are expected to be the main hardware limitation \cite{Bjornson2014a} and henceforth we only consider those impairments in this section.

\begin{figure*}[!t]
\normalsize

\begin{equation} \label{eq:SINR-value-impairment} \tag{35}
\widetilde{\mathrm{SINR}}_{jk} =  \frac{ (1-\epsilon^2) p_{jk} | \mathbb{E}_{\{\vect{h}\}}\{ \vect{g}_{jk}^{\Htran} \vect{h}_{jjk} \} |^2 }{ \fracSum{l \in \mathcal{L}} \fracSumtwo{m=1}{K} p_{lm}  \mathbb{E}_{\{\vect{h}\}} \{ |\vect{g}_{jk}^{\Htran} \vect{h}_{jlm} |^2  \} - (1-\epsilon^2)  p_{jk} | \mathbb{E}_{\{\vect{h}\}} \{ \vect{g}_{jk}^{\Htran} \vect{h}_{jjk} \} |^2  + \sigma^2  \mathbb{E}_{\{\vect{h}\}} \{ \| \vect{g}_{jk} \|^2\}  }.
\end{equation}
\hrulefill
\end{figure*}

Similar to \cite{Schenk2008a,wenk2010mimo,Zhang2012a}, we model the hardware impairments as a reduction of the original signals by a factor $\sqrt{1-\epsilon^2}$ and replacing it with Gaussian distortion noise that carries the removed power. More precisely, the UL system model in \eqref{eq:system-model} is generalized as
\begin{equation} \label{eq:system-model-impair}
\vect{y}_j = \sum_{l \in \mathcal{L}} \sum_{k=1}^{K} \vect{h}_{jlk} \left( \sqrt{ (1-\epsilon^2) p_{lk} } x_{lk}  + \varepsilon_{lk} \right) + \vect{n}_{j},
\end{equation}
where $\varepsilon_{lk} \sim \mathcal{CN}(0, \epsilon^2 p_{lk}  )$ is the UL distortion noise caused at UE $k$ in cell $l$, and
the DL system model in \eqref{eq:system-model-DL} is generalized as
\begin{equation} \label{eq:system-model-DL-impair}
z_{jk} = \sqrt{1-\epsilon^2} \left(\sum_{l \in \mathcal{L}} \sum_{m=1}^{K} \vect{h}_{ljk}^{\Ttran} \vect{w}_{lm} s_{lm} + \eta_{jk} \right) + e_{jk},
\end{equation}
where $e_{jk} \sim \mathcal{CN}(0, \epsilon^2 ( \sum_{l \in \mathcal{L}} \sum_{m=1}^{K} \| \vect{h}_{ljk}^{\Ttran} \vect{w}_{lm} \|^2 + \sigma^2  ) )$ is the DL distortion noise caused at UE $k$ in cell $j$. Notice that $\sum_{l \in \mathcal{L}} \sum_{m=1}^{K} \| \vect{h}_{ljk}^{\Ttran} \vect{w}_{lm} \|^2 + \sigma^2$ is the power of the term in parenthesis in  \eqref{eq:system-model-DL-impair}. The parameter $\epsilon$ determines the level of impairments and can be interpreted as the error vector magnitude (EVM) \cite{wenk2010mimo}; typical values in LTE are in the range $0 \leq \epsilon \leq 0.17$ \cite{Holma2011a}. Based on these generalized system models, the following counterpart of Lemmas \ref{lemma:SE} and \ref{lemma:SE-DL} is obtained.

\begin{lemma} \label{lemma:SE-impairment}
Under hardware impairments, a jointly achievable SE in the UL and DL of {an arbitrary} UE $k$ in cell $j$ is
\begin{equation}
\left( 1-\frac{B}{S} \right) \mathbb{E}_{\{\vect{z}\}} \left\{ \log_2(1+ \widetilde{\mathrm{SINR}}_{jk}) \right\} \quad \text{[bit/s/Hz]}
\end{equation}
where the effective SINR is given in \eqref{eq:SINR-value-impairment} at the top of the page. \setcounter{equation}{35}
\end{lemma}
\begin{IEEEproof}
The proof is given in the appendix.
\end{IEEEproof}

The SE expression in Lemma \ref{lemma:SE-impairment} resembles our previous results in Section \ref{sec:average-SE}, with the only differences that there is a loss in desired signal power by a factor $(1-\epsilon^2)$ and that this power is turned into self-interference in the denominator of the SINR. Under the assumption of MR, ZF, or P-ZF processing in the UL and DL, we have the following closed-form SE expression.

\begin{theorem} \label{theorem-achievable-rate-impairments}
 Let $\mathcal{L}_j(\beta) \subset \mathcal{L}$ be the subset of cells that uses the same pilots as cell $j$. Looking jointly at the UL and DL, an achievable SE in cell $j$ under hardware impairments is \vskip-5mm
\begin{equation}
\mathrm{SE}_j = K \left( 1-\frac{B}{S} \right) \log_2 \left(1+ \frac{1-\epsilon^2}{ I_{j}^{\mathrm{scheme}} +  \epsilon^2 } \right) \, \text{[bit/s/Hz/cell]}
\end{equation}
where the interference term $I_{j}^{\mathrm{scheme}}$ is defined in \eqref{eq:achievable-SINR-ULscheme} and depends on $G^{\mathrm{scheme}}$ and $Z_{jl}^{\mathrm{scheme}}$. The parameter values with MR, ZF, and P-ZF are as follows:

\begin{center}
  \begin{tabular}{ | c | c | c | }
    \hline
    Scheme & $G^{\mathrm{scheme}}$ & $Z_{jl}^{\mathrm{scheme}}$ \\ \hline
    MR & $\!M (1-\epsilon^2)\!$ & $K$ \\ \hline
    ZF & $\!(M-K) (1-\epsilon^2)\!$ & 
    $ \begin{cases}
      K\left( 1-
\frac{ (1-\epsilon^2) \mu_{j l}^{(1)}  }{  \fracSum{\ell \in \mathcal{L}_j(\beta) }  \mu_{j \ell}^{(1)} + \frac{\sigma^2}{B  \rho} }
 \right) \!\!\!\!\!\!\!\!\!\! \\ \quad \textrm{if } l  \in  \mathcal{L}_j(\beta) \\
    K \,\, \textrm{if } l \not \in  \mathcal{L}_j(\beta)
    \end{cases}$ \\ \hline
    P-ZF & $\!(M-B) (1-\epsilon^2)\!$ & $
    K   \left( \!
1-
\frac{ (1-\epsilon^2) \mu_{j l}^{(1)} }{  \fracSum{\ell \in \mathcal{L}_l(\beta) } \mu_{j \ell}^{(1)}   + \frac{\sigma^2}{ B \rho} }
\! \right)$ \\
    \hline
  \end{tabular}
\end{center}

If $M \rightarrow \infty$ (with $K,B \leq S < \infty$), the effective SINRs with these processing schemes approach the upper limit
\begin{equation} \label{eq:asymptotic-SINR-impairment}
 \frac{ 1-\epsilon^2 }{ \fracSum{l \in \mathcal{L}_j(\beta) \setminus \{ j \} } \mu^{(2)}_{jl}  + \epsilon^2  }.
\end{equation}
\end{theorem}
\begin{IEEEproof}
This result follows straightforwardly from Theorems \ref{theorem-achievable-rate-conventional}--\ref{theorem:DLUL-duality}, since the SINR expressions in \eqref{eq:SINR-value-impairment} only differ from those in Section \ref{sec:average-SE} by the $(1-\epsilon^2)$-factors.
\end{IEEEproof}

 \begin{figure}[!t]
\begin{center} 
\includegraphics[width=\columnwidth]{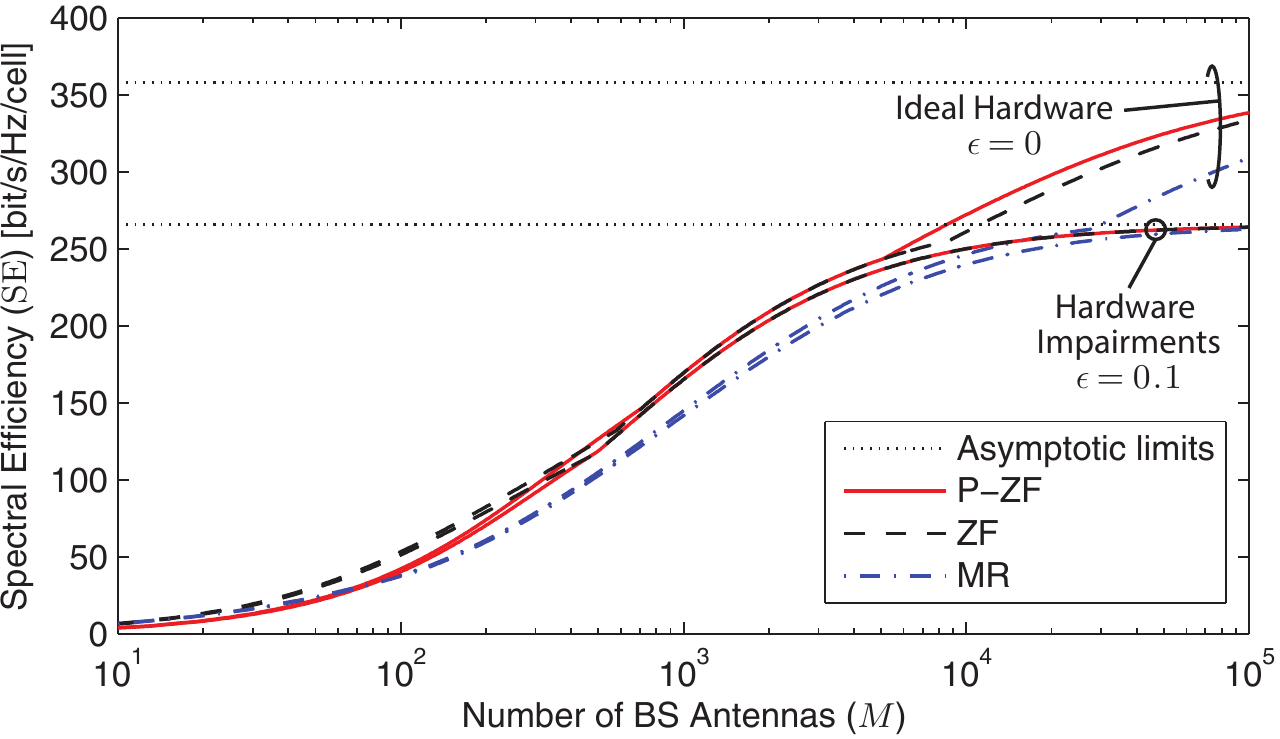}
\end{center}\vskip-3mm
\caption{Optimized per-cell SE with or without hardware impairments.} \label{figure_journal_SE_mean_impairments} \vskip-2mm
\end{figure}

Using the tractable SE expression in Theorem \ref{theorem-achievable-rate-impairments} for simulation, Fig.~\ref{figure_journal_SE_mean_impairments} shows the per-cell SE in the average inter-cell interference. This figure shows results for ideal hardware with $\epsilon = 0$ (as in Fig.~\ref{figure_simulation_mean}(a)) and for hardware impairments with $\epsilon = 0.1$, which is a large EVM number in these contexts \cite{Holma2011a}. Interestingly, there is only a tiny difference in SE for $M < 5000$, mainly because the SE per UE is relatively small at the optimized operating points and thus the distortion noise is only a minor limiting factor. For higher number of antennas, the difference is substantial because of the asymptotic limits for ideal hardware in \eqref{eq:asymptotic-SINR} and for hardware impairments in \eqref{eq:asymptotic-SINR-impairment} are different. We conclude that hardware impairments seem to have a small impact on practical massive MIMO systems, which have been optimized for high SE.

\begin{figure*}[!t]
\normalsize

\begin{align} \tag{42} \label{eq:SINR-UL-MR}
\mathrm{SINR}_{jk}^{\rm{(ul)}}  = \frac{ \vect{v}_{i_{jk}}^{\Htran} \boldsymbol{\Psi}^{-1}_j \vect{v}_{i_{jk}}  }{ \fracSum{l \in \mathcal{L}} \fracSumtwo{m=1}{K} \left(
\frac{d_j(\vect{z}_{lm}) }{ d_l(\vect{z}_{lm})} \frac{1}{M}  +
\left( \frac{d_j(\vect{z}_{lm}) }{ d_l(\vect{z}_{lm})} \right)^2 \vect{v}_{i_{jk}}^{\Htran} \boldsymbol{\Psi}^{-1}_j \vect{v}_{i_{lm}} \right) -  \vect{v}_{i_{jk}}^{\Htran} \boldsymbol{\Psi}^{-1}_j \vect{v}_{i_{jk}}   + \frac{\sigma^2}{  M \rho}   }
\end{align}
\hrulefill
\end{figure*}

\section{Conclusion}
\label{sec:conclusion}

This paper investigated how many UEs, $K$, that should be scheduled in massive MIMO systems to maximize the SE per cell for a fixed $M$. Conventional SE expressions are strongly dependent on the UE positions, which makes it hard to optimize $K$. In contrast, we derived new SE expressions that are independent of the instantaneous UE positions, due to power control and averaging over random UE locations. In fact, the new expressions are the same for the UL and DL, which allows for joint network optimization. When applied to symmetric network topologies, where each cell is representable for any cell, these expressions can directly provide the network-wide performance---which otherwise would require extensive Monte-Carlo simulations. 

The analytic results treat MR and ZF processing and a new distributed cooperation scheme, coined P-ZF, that suppresses inter-cell interference by listening to the pilot transmissions from neighboring cells. The asymptotic analysis shows that the SE-optimal $K^*$ approaches $\frac{S}{2\beta}$ as $M \rightarrow \infty$, irrespective of the processing scheme. Hence, $B = \beta K^* \rightarrow \frac{S}{2}$ which means that half the frame should be spent on pilot signaling when $M$ is large enough. The corresponding asymptotic SE limit is not reached for practical $M$, but an unconventionally large fraction of the frame should still be allocated to pilots: 5\% to 40\% were observed in simulations for $M \leq 1000$.

Generally speaking, high per-cell SEs are achieved by scheduling many UEs for simultaneous transmission, while the SE per UE might only be 1--4 bit/s/Hz. P-ZF gives the highest performance per UE, while MR gives the lowest SE per UE. In contrast, MR schedules the largest number of UEs and P-ZF the smallest number. ZF processing is often the best choice in terms of per-cell SE, thus showing that the inter-cell interference suppression offered by P-ZF is only needed in special cases with strong inter-cell interference. The extensive simulations show that massive MIMO prefers an SNR of 0--5 dB, that a non-universal pilot reuse of $\beta = 3$ is often a decent choice, and that the technology is very robust to distortion noise from hardware impairments. Based on the simulations, we notice that massive MIMO with $M=100$ can easily achieve a $10\times$ gain in SE over the IMT-Advanced requirement of 3 bit/s/Hz/cell. For large arrays with $M=500$ antennas, massive MIMO can even provide a $40 \times $ gain over IMT-Advanced. {The results in this paper are for uncorrelated fading, while spatially correlated fading is expected to reduce the inter-user interference \cite{Yin2013a} thus leading to higher SEs and allowing for smaller $\beta$.}

\appendices

\section*{Appendix: Collection of Proofs}

\textbf{Proof of Lemma \ref{lemma:LMMSE-estimation}:} As shown in \cite[Theorem 11.1]{Kay1993a}, the expression for an MMSE estimator $\hat{\vect{h}}^{\mathrm{eff}}_{jlk}$ of $\vect{h}^{\mathrm{eff}}_{jlk}$ with jointly Gaussian channels and (colored) noise is
\begin{equation} \label{eq:MMSE-expression}
\mathbb{E}_{\{ \vect{h} \}} \{ \vect{h}^{\mathrm{eff}}_{jlk}  \vectorize(\vect{Y}_j)^{\Htran} \}  \left( \mathbb{E}_{\{ \vect{h} \}} \{  \vectorize(\vect{Y}_j) \vectorize(\vect{Y}_j)^{\Htran} \}  \right)^{-1}   \vectorize(\vect{Y}_j)
\end{equation}
where $\vectorize(\cdot)$ denotes vectorization. Direct algebraic computation using the vectorization rule  $(\vect{C}^{\Ttran} \kron \vect{A}) \vectorize(\vect{B}) = \vectorize(\vect{A} \vect{B} \vect{C})$, where $\kron$ is the Kronecker product, shows that
\begin{align} \notag
\mathbb{E}_{\{ \vect{h} \}} \{ \vect{h}^{\mathrm{eff}}_{jlk}  \vectorize(\vect{Y}_j)^{\Htran} \} &= \mathbb{E}_{\{ \vect{h} \}} \left\{ \vect{h}^{\mathrm{eff}}_{jlk} (\vect{h}^{\mathrm{eff}}_{jlk})^{\Htran}
\left( \vect{v}_{i_{lk}}^{\Htran}  \kron \vect{I}_M  \right) \right\} \\ &= \left( \vect{v}_{i_{lk}}^{\Htran}  \kron \rho \frac{d_j(\vect{z}_{lk}) }{ d_l(\vect{z}_{lk})}  \vect{I}_M  \right) \label{eq:MMSE-firstexpectation}
\end{align}
since the channels are independent. Similarly, the mutual independence of the UE channels implies that
\begin{align} \notag
& \mathbb{E}_{\{ \vect{h} \}} \{  \vectorize(\vect{Y}_j) \vectorize(\vect{Y}_j)^{\Htran} \} = \sigma^2 \vect{I}_{MB} \\  \notag &+  \sum_{ \ell \in \mathcal{L}} \sum_{m=1}^{K}
\mathbb{E}_{\{ \vect{h}_{j \ell m} \}} \left\{ \vectorize( \vect{h}^{\mathrm{eff}}_{j\ell m} \vect{v}_{i_{\ell m}}^{\Ttran} ) \vectorize( \vect{h}^{\mathrm{eff}}_{j\ell m} \vect{v}_{i_{\ell m}}^{\Ttran} )^{\Htran} \right\}  \\ &= \left( \sum_{ \ell \in \mathcal{L}} \sum_{m=1}^{K} \rho \frac{d_j(\vect{z}_{\ell m}) }{ d_{\ell}(\vect{z}_{\ell m})}  \vect{v}_{i_{\ell m}} \vect{v}_{i_{\ell m}}^{\Htran} + \sigma^2 \vect{I}_B  \right) \kron \vect{I}_M. \label{eq:MMSE-secondexpectation}
\end{align}

The expression \eqref{eq:LMMSE-estimator} is obtained by substituting \eqref{eq:MMSE-firstexpectation} and \eqref{eq:MMSE-secondexpectation} into \eqref{eq:MMSE-expression}, normalizing by $\rho$ and using the vectorization rule above. According to the definition in \eqref{eq:LMMSE-error-cov}, the error covariance matrix $\vect{C}_{jlk}$ is given by
\begin{equation} \label{eq:LMMSE-error-cov-computation}
\begin{split}
&\rho \frac{d_j(\vect{z}_{lk}) }{ d_l(\vect{z}_{lk})}  \vect{I}_M - \mathbb{E}_{\{ \vect{h} \}} \{ \vect{h}^{\mathrm{eff}}_{jlk}  \vectorize(\vect{Y}_j)^{\Htran} \} \\ & \times \left( \mathbb{E}_{\{ \vect{h} \}} \{  \vectorize(\vect{Y}_j) \vectorize(\vect{Y}_j)^{\Htran} \}  \right)^{-1} \mathbb{E}_{\{ \vect{h} \}} \{ \vect{h}^{\mathrm{eff}}_{jlk}  \vectorize(\vect{Y}_j)^{\Htran} \}^{\Htran} \\
& = \rho \frac{d_j(\vect{z}_{lk}) }{ d_l(\vect{z}_{lk})} \left( 1 - \frac{d_j(\vect{z}_{lk}) }{ d_l(\vect{z}_{lk})} \vect{v}_{i_{lk}}^{\Htran} \boldsymbol{\Psi}^{-1}_j \vect{v}_{i_{lk}} \right) \vect{I}_M \\ &= \rho \frac{d_j(\vect{z}_{lk}) }{ d_l(\vect{z}_{lk})} \left( 1 - \frac{\frac{d_j(\vect{z}_{lk}) }{ d_l(\vect{z}_{lk})} B}{\sum_{\ell \in \mathcal{L}} \sum_{m=1}^{K} \frac{d_j(\vect{z}_{\ell m}) }{ d_{\ell}(\vect{z}_{\ell m})} \vect{v}_{i_{lk}}^{\Htran} \vect{v}_{i_{\ell m}} + \frac{\sigma^2}{\rho}} \right) \vect{I}_M,
\end{split}
\end{equation}
where the last equality follows from the fact that the pilot signals form an orthogonal basis.

\vspace{3mm}

\begin{figure*}[!t]
\normalsize

\begin{align} \tag{49} \label{eq:SINR-UL-ZF-derivation}
\mathrm{SINR}_{jk}^{\rm{(ul)}}  = \frac{ 1 }{ \fracSum{l \in \mathcal{L}} \fracSumtwo{m=1}{K}
\left(\frac{d_j(\vect{z}_{lm}) }{ d_l(\vect{z}_{lm})} \right)^2 \frac{\vect{v}_{i_{jk}}^{\Htran}\vect{v}_{i_{l m}} }{B} +
\frac{\rho \frac{d_j(\vect{z}_{lm}) }{ d_l(\vect{z}_{lm})} \left( 1 - A^{\mathrm{ZF}}_{j l} \frac{d_j(\vect{z}_{lm}) }{ d_l(\vect{z}_{lm})} \vect{v}_{i_{lm}}^{\Htran} \boldsymbol{\Psi}^{-1}_j \vect{v}_{i_{lm}} \right)}{(M-K) \rho \vect{v}_{i_{jk}}^{\Htran} \boldsymbol{\Psi}^{-1}_j \vect{v}_{i_{jk}}  } - 1  +   \frac{\sigma^2}{(M-K) \rho \vect{v}_{i_{jk}}^{\Htran} \boldsymbol{\Psi}^{-1}_j \vect{v}_{i_{jk}}}  }
\end{align}
\begin{align} \tag{54} \label{eq:SINR-UL-PZF-derivation}
\mathrm{SINR}_{jk}^{\rm{(ul)}}  = \frac{ 1 }{ \fracSum{l \in \mathcal{L}} \fracSumtwo{m=1}{K}
\left(\frac{d_j(\vect{z}_{lm}) }{ d_l(\vect{z}_{lm})} \right)^2 \frac{\vect{v}_{i_{jk}}^{\Htran}\vect{v}_{i_{l m}} }{B} +
\frac{\rho \frac{d_j(\vect{z}_{lm}) }{ d_l(\vect{z}_{lm})} \left( 1 - \frac{d_j(\vect{z}_{lm}) }{ d_l(\vect{z}_{lm})} \vect{v}_{i_{lm}}^{\Htran} \boldsymbol{\Psi}^{-1}_j \vect{v}_{i_{lm}} \right)}{(M-B) \rho \vect{v}_{i_{jk}}^{\Htran} \boldsymbol{\Psi}^{-1}_j \vect{v}_{i_{jk}}  } - 1  +   \frac{\sigma^2}{(M-B) \rho \vect{v}_{i_{jk}}^{\Htran} \boldsymbol{\Psi}^{-1}_j \vect{v}_{i_{jk}}}  }
\end{align}
\hrulefill
\end{figure*}

\textbf{Proof of Theorem \ref{theorem-achievable-rate-conventional}:} The first step for MR combining is to compute the expectations in \eqref{eq:SINR-value} with respect to the channel realizations.
These are obtained from \cite[Corollary 2]{Bjornson2015b} by setting $\kappa=\delta =0$, $\xi = \sigma^2$, and $\lambda_{jlm} = \frac{d_j(\vect{z}_{lm}) }{ d_l(\vect{z}_{lm})}$.
Plugging these expressions into \eqref{eq:SINR-value} yields, for MR, the expression in \eqref{eq:SINR-UL-MR} at the top of the page, by multiplying each term by $\frac{1}{ M^2 \rho^2 \vect{v}_{i_{jk}}^{\Htran} \boldsymbol{\Psi}^{-1}_j \vect{v}_{i_{jk}}}$. \setcounter{equation}{42}
The expression in {\eqref{eq:achievable-SINR-ULscheme} for MR} is now obtained by considering an achievable lower bound $\mathbb{E}_{\{\vect{z}\}} \{\log_2(1+ \frac{1}{f(\{\vect{z}\})}) \} \geq \log_2(1+ \frac{1}{\mathbb{E}_{\{\vect{z}\}} \{ f(\{\vect{z}\}) \} }) $ where the expectation with respect to user positions is moved to the denominator of the SINRs using Jensen's inequality. This leads to expectations of the following types:
\begin{align} \notag 
&\mathbb{E}_{\{\vect{z}\}} \left\{ \frac{1}{ \vect{v}_{i_{jk}}^{\Htran} \boldsymbol{\Psi}^{-1}_j \vect{v}_{i_{jk}} } \right\} \\  \notag &=
\mathbb{E}_{\{\vect{z}\}} \left\{ \frac{\sum_{\ell \in \mathcal{L}} \sum_{\tilde{m}=1}^{K} \frac{d_j(\vect{z}_{\ell \tilde{m}}) }{ d_{\ell}(\vect{z}_{\ell \tilde{m}})} \vect{v}_{i_{jk}}^{\Htran} \vect{v}_{i_{\ell \tilde{m}}} + \frac{\sigma^2}{\rho} }{B} \right\}  
\\ \label{eq:expectation-wrt-z}
&=  \frac{ \sum_{\ell \in \mathcal{L}_j (\beta) } \mu_{j \ell}^{(1)} B + \frac{\sigma^2}{\rho} }{B} 
\end{align}
\begin{align}
\mathbb{E}_{\{\vect{z}\}} \left\{  \fracSum{l \in \mathcal{L}} \fracSumtwo{m=1}{K}
\left( \frac{d_j(\vect{z}_{lm}) }{ d_l(\vect{z}_{lm})} \right)^2 \frac{\vect{v}_{i_{jk}}^{\Htran} \boldsymbol{\Psi}^{-1}_j \vect{v}_{i_{lm}}}{ \vect{v}_{i_{jk}}^{\Htran} \boldsymbol{\Psi}^{-1}_j \vect{v}_{i_{jk}}}   \right\} &  = \fracSum{l \in \mathcal{L}_j (\beta)}   \mu_{j l}^{(2)} 
\end{align}
\begin{align} \notag
& \mathbb{E}_{\{\vect{z}\}} \left\{  \fracSum{l \in \mathcal{L}} \fracSumtwo{m=1}{K} \frac{\frac{d_j(\vect{z}_{lm}) }{ d_l(\vect{z}_{lm})}}{ \vect{v}_{i_{jk}}^{\Htran} \boldsymbol{\Psi}^{-1}_j \vect{v}_{i_{jk}} } \right\} \\ &=  \fracSum{l \in \mathcal{L}} K
\mu_{j l}^{(1)} \frac{ \sum_{\ell \in \mathcal{L}_j (\beta)  } \mu_{j \ell}^{(1)} B + \frac{\sigma^2}{\rho} }{B}
+ \sum_{l \in \mathcal{L}_j (\beta)} \mu_{j l}^{(2)}-( \mu_{j l}^{(1)} )^2 \label{eq:expectation-wrt-z2}
\end{align}
where we have utilized the definition in \eqref{eq:mu-definition} {and the non-universal pilot reuse assumption} to identify the expectations.

The expectations in \eqref{eq:SINR-value} with respect to the channel realizations for ZF combining are
\begin{align} \label{eq:ZF1}
 \mathbb{E}_{\{ \vect{h}\}} \{ \| \vect{g}_{jk}^{\mathrm{ZF}} \|^2\} & = \frac{1}{(M-K) \rho \vect{v}_{i_{jk}}^{\Htran} \boldsymbol{\Psi}^{-1}_j \vect{v}_{i_{jk}}  } \\ \label{eq:ZF2}
p_{jk} | \mathbb{E}_{\{ \vect{h}\}} \{ (\vect{g}_{jk}^{\mathrm{ZF}})^{\Htran} \vect{h}_{jjk} \} |^2 &= 1
\end{align}
\begin{align} \notag
&p_{lm}  \mathbb{E}_{\{ \vect{h}\}} \{ | (\vect{g}_{jk}^{\mathrm{ZF}})^{\Htran} \vect{h}_{jlm} |^2  \} = \left(\frac{d_j(\vect{z}_{lm}) }{ d_l(\vect{z}_{lm})} \right)^2 \frac{\vect{v}_{i_{jk}}^{\Htran}\vect{v}_{i_{l m}} }{B}  \\ &+
\frac{\rho \frac{d_j(\vect{z}_{lm}) }{ d_l(\vect{z}_{lm})} \left( 1 - A^{\mathrm{ZF}}_{j l} \frac{d_j(\vect{z}_{lm}) }{ d_l(\vect{z}_{lm})} \vect{v}_{i_{lm}}^{\Htran} \boldsymbol{\Psi}^{-1}_j \vect{v}_{i_{lm}} \right)}{(M-K) \rho \vect{v}_{i_{jk}}^{\Htran} \boldsymbol{\Psi}^{-1}_j \vect{v}_{i_{jk}}  } \label{eq:ZF3}
\end{align}
 with $A^{\mathrm{ZF}}_{j l} = 1$ if $l \in \mathcal{L}_j(\beta)$ and zero otherwise, where \eqref{eq:ZF1} follows from the definition of ZF and by utilizing well-known properties of Wishart matrices (see e.g.,
\cite[Proof of Proposition 2]{Ngo2013a}) and \eqref{eq:ZF2} is a consequence of the ZF principle.
The first term in \eqref{eq:ZF3} follows from \eqref{eq:ZF2} whenever $\vect{v}_{i_{jk}}^{\Htran}\vect{v}_{i_{l m}} \neq 0$ (i.e., when the same pilot signal is used). The second term is the product between $\mathbb{E}_{\{ \vect{h}\}}\{ \| \vect{g}_{jk}^{\mathrm{ZF}} \|^2\}$ and the variance of the estimation error of the effective channel $\sqrt{p_{lm}} \vect{h}_{jlm}$ if $A^{\mathrm{ZF}}_{j l} \neq 0$ (i.e., {if the UE is in a cell $l \in \mathcal{L}_j (\beta)$}) or the original variance of $\sqrt{p_{lm}} \vect{h}_{jlm}$ if $A^{\mathrm{ZF}}_{j l}=0$. \setcounter{equation}{49}
Using \eqref{eq:ZF1}--\eqref{eq:ZF3}, we obtain the expression \eqref{eq:SINR-UL-ZF-derivation} at the of the page for ZF. Finally, the achievable SE in the theorem is obtained by using Jensen's inequality in the same way as for MR, where the expectation in \eqref{eq:expectation-wrt-z}--\eqref{eq:expectation-wrt-z2} reappear along with
\begin{equation} 
\begin{split}
&\mathbb{E}_{\{\vect{z}\}} \left\{ -
\left( \frac{d_j(\vect{z}_{lm}) }{ d_l(\vect{z}_{lm})} \right)^2 \frac{ \vect{v}_{i_{lm}}^{\Htran} \boldsymbol{\Psi}^{-1}_j \vect{v}_{i_{lm}} }{\vect{v}_{i_{jk}}^{\Htran} \boldsymbol{\Psi}^{-1}_j \vect{v}_{i_{jk}} } \right\} \\ &\leq - (\mu_{jl}^{(1)})^2
\frac{  \fracSum{\ell \in \mathcal{L}_j (\beta) } \mu_{j \ell}^{(1)}  + \frac{\sigma^2}{ B \rho}  }{  \fracSum{\ell \in \mathcal{L}_l (\beta) } \mu_{j \ell}^{(1)}  + \frac{\sigma^2}{ B \rho} }, 
\end{split}
\end{equation}
where the inequality is once again from Jensen's inequality.

\vspace{3mm}

\textbf{Proof of Theorem \ref{theorem-achievable-rate-pilotbased}:} Similar to the proof of Theorem \ref{theorem-achievable-rate-conventional}, for P-ZF we obtain
\begin{align} \label{eq:PZF1}
 \mathbb{E}_{\{ \vect{h}\}} \{ \| \vect{g}_{jk}^{\mathrm{P}\text{-}\mathrm{ZF}} \|^2\} & = \frac{1}{(M-B) \rho \vect{v}_{i_{jk}}^{\Htran} \boldsymbol{\Psi}^{-1}_j \vect{v}_{i_{jk}}  } 
 \end{align}
 \begin{align}
  \label{eq:PZF2}
p_{jk} | \mathbb{E}_{\{ \vect{h}\}} \{ (\vect{g}_{jk}^{\mathrm{P}\text{-}\mathrm{ZF}})^{\Htran} \vect{h}_{jjk} \} |^2 &= 1
\end{align}
\begin{align} \notag
&p_{lm}  \mathbb{E}_{\{ \vect{h}\}} \{ | (\vect{g}_{jk}^{\mathrm{P}\text{-}\mathrm{ZF}})^{\Htran} \vect{h}_{jlm} |^2  \} = \left(\frac{d_j(\vect{z}_{lm}) }{ d_l(\vect{z}_{lm})} \right)^2 \frac{\vect{v}_{i_{jk}}^{\Htran}\vect{v}_{i_{l m}} }{B}  \\ &+
\frac{\rho \frac{d_j(\vect{z}_{lm}) }{ d_l(\vect{z}_{lm})} \left( 1 - \frac{d_j(\vect{z}_{lm}) }{ d_l(\vect{z}_{lm})} \vect{v}_{i_{lm}}^{\Htran} \boldsymbol{\Psi}^{-1}_j \vect{v}_{i_{lm}} \right)}{(M-B) \rho \vect{v}_{i_{jk}}^{\Htran} \boldsymbol{\Psi}^{-1}_j \vect{v}_{i_{jk}}  } \label{eq:PZF3}
\end{align}
by following the procedures used for ZF. \setcounter{equation}{54}
Using \eqref{eq:PZF1}--\eqref{eq:PZF3}, we obtain the expression \eqref{eq:SINR-UL-PZF-derivation} at the top of the page for P-ZF. The final expression is obtained by considering an achievable lower bound $\mathbb{E}_{\{\vect{z}\}} \{\log_2(1+ \frac{1}{f(\vect{z})}) \} \geq \log_2(1+ \frac{1}{\mathbb{E}_{\{\vect{z}\}} \{ f(\vect{z}) \} }) $ using Jensen's inequality, similar to the ZF case in Theorem\ref{theorem-achievable-rate-conventional}.

\vspace{3mm}

\begin{figure*}[!t]
\normalsize
\begin{equation} \tag{61} \label{eq:lemma-impairment-derivation}
\begin{split}
\frac{ (1-\epsilon^2)  |\mathbb{E}\{ s \}|^2}{ (1-\epsilon^2) \left( \mathbb{E}\{ |n|^2 \} + \mathbb{E}\{ |s|^2 \} - |\mathbb{E}\{ s \}|^2 \right) + \epsilon^2 \left( \mathbb{E}\{ |n|^2 \} + \mathbb{E}\{ |s|^2 \}   \right)   }  = \frac{ (1-\epsilon^2)  |\mathbb{E}\{ s \}|^2}{ \mathbb{E}\{ |n|^2 \} + \mathbb{E}\{ |s|^2 \} - (1-\epsilon^2) |\mathbb{E}\{ s \}|^2    }
\end{split}
\end{equation}
\hrulefill
\end{figure*}

\textbf{Proof of Theorem \ref{theorem:DLUL-duality}:} Suppose that $\gamma_{jk} = \mathrm{SINR}_{jk}^{\rm{(ul)}}$ is the UL SINR value achieved by UE $k$ in cell $j$ for a given receive combining scheme. The goal of the proof is to show that we can also achieve $\gamma_{jk} = \mathrm{SINR}_{jk}^{\rm{(dl)}} $ for the DL SINR in \eqref{eq:SINR-value-DL}. This condition can also be expressed as
\begin{equation} \label{eq:SINR-condition-DL}
\begin{split}
&\frac{\gamma_{jk} \mathbb{E}_{\{\vect{h}\}} \{\| \check{\vect{g}}_{jk} \|^2 \} }{ | \mathbb{E}_{\{\vect{h}\}}\{ \check{\vect{g}}_{jk}^{\Htran} \vect{h}_{jjk} \} |^2 } \\ &=  \frac{ q_{jk}  }{ \fracSum{l \in \mathcal{L}} \fracSumtwo{m=1}{K} q_{lm} \frac{ \mathbb{E}_{\{\vect{h}\}} \{ |\check{\vect{g}}_{lm}^{\Htran} \vect{h}_{ljk} |^2  \} }{ \mathbb{E}_{\{\vect{h}\}} \{\| \check{\vect{g}}_{lm} \|^2 \} }- q_{jk} \frac{| \mathbb{E}_{\{\vect{h}\}}\{ \check{\vect{g}}_{jk}^{\Htran} \vect{h}_{jjk} \} |^2}{ \mathbb{E}_{\{\vect{h}\}} \{\| \check{\vect{g}}_{jk} \|^2 \}}  + \sigma^2 }.
\end{split}
\end{equation}
We define the $K |\mathcal{L}| \times K |\mathcal{L}|$ block matrix $\boldsymbol{\Psi}$, where each block is $K \times K$ and the $(j,l)$th block is denoted $\boldsymbol{\Psi}_{jl}$. Its $(k,m)$th element is given by
\begin{equation}
\begin{split}
&[\boldsymbol{\Psi}_{jl}]_{k,m} \\ &= \begin{cases}  \frac{ \mathbb{E}_{\{\vect{h}\}} \{ |\check{\vect{g}}_{lm}^{\Htran} \vect{h}_{ljk} |^2  \} }{ \mathbb{E}_{\{\vect{h}\}} \{\| \check{\vect{g}}_{lm} \|^2 \} } - \frac{| \mathbb{E}_{\{\vect{h}\}}\{ \check{\vect{g}}_{jk}^{\Htran} \vect{h}_{jjk} \} |^2}{ \mathbb{E}_{\{\vect{h}\}} \{\| \check{\vect{g}}_{jk} \|^2 \}} & \textrm{if } k=m, j=l, \\
\frac{ \mathbb{E}_{\{\vect{h}\}} \{ |\check{\vect{g}}_{lm}^{\Htran} \vect{h}_{ljk} |^2  \} }{ \mathbb{E}_{\{\vect{h}\}} \{\| \check{\vect{g}}_{lm} \|^2 \} } & \textrm{otherwise}.
 \end{cases}
 \end{split}
\end{equation}
Moreover, we define the $K |\mathcal{L}| \times K |\mathcal{L}|$ block diagonal matrix $\vect{D}$, where the $j$th $K \times K$ block is $\vect{D}_{j}$ and its $k$th diagonal element is
\begin{equation}
[ \vect{D}_{j} ]_{k,k} = \frac{\gamma_{jk} \mathbb{E}_{\{\vect{h}\}} \{\| \check{\vect{g}}_{jk} \|^2 \} }{ | \mathbb{E}_{\{\vect{h}\}}\{ \check{\vect{g}}_{jk}^{\Htran} \vect{h}_{jjk} \} |^2 } .
\end{equation}
Using this notation, \eqref{eq:SINR-condition-DL} can be expressed as
\begin{equation} \label{eq:SINR-condition-DL2}
\begin{split}
&[ \vect{D}_{j} ]_{k,k}  =  \frac{ q_{jk}  }{ \fracSum{l \in \mathcal{L}} \fracSumtwo{m=1}{K} q_{lm} [\boldsymbol{\Psi}_{jl}]_{k,m}  + \sigma^2 } \\ & \Leftrightarrow \quad
[ \vect{D}_{j} ]_{k,k} \sigma^2  =   q_{jk} - \fracSum{l \in \mathcal{L}} \fracSumtwo{m=1}{K} q_{lm} [ \vect{D}_{j} ]_{k,k} [\boldsymbol{\Psi}_{jl}]_{k,m}.
\end{split}
\end{equation}
In matrix form, the DL SINR conditions for all UEs in all cells can be expressed as
$\vect{D} \sigma^2 = \vect{q} - \vect{D} \boldsymbol{\Psi} \vect{q}$,
where $\vect{q} = [ \vect{q}_1^{\Ttran} \, \ldots \, \vect{q}_{|\mathcal{L}|}^{\Ttran}]^{\Ttran}$ and $\vect{q}_j = [q_{j1} \, \ldots \, q_{jK}]^{\Ttran}$ contain the DL transmit powers in the $j$th cell.
This expression can now be solved for $\vect{q}$.
The matrix $\vect{D}$ depends only on the precoding vectors, thus for any choice of precoding scheme the sought SINRs are achieved by the power control policy
\begin{equation}  \label{eq:DL-power}
\vect{q}^{\star}  = \sigma^2 (\vect{I}_{K |\mathcal{L}|} - \vect{D} \boldsymbol{\Psi} )^{-1}\vect{D} \vect{1}
\end{equation}
where $\vect{1}$ is the vector with only ones. $\vect{q}^{\star}$ is a feasible power control (i.e., has positive values) if all eigenvalues of $(\vect{I}_{K |\mathcal{L}|} - \vect{D} \boldsymbol{\Psi} )$ are larger than zero. We need to show that this always holds. We notice that the UL SINR condition, which is satisfied by assumption, can be expressed in a similar matrix form where $\boldsymbol{\Psi}$ is replaced by $\boldsymbol{\Psi}^{\Ttran}$:
\begin{equation} \label{eq:condition-power2}
\vect{D} \sigma^2 = \vect{p} - \vect{D} \boldsymbol{\Psi}^{\Ttran} \vect{p} \quad \Leftrightarrow \quad \vect{p} = \sigma^2 (\vect{I}_{K |\mathcal{L}|} - \vect{D} \boldsymbol{\Psi}^{\Ttran} )^{-1}\vect{D} \vect{1},
\end{equation}
where $\vect{p} = [ \vect{p}_1^{\Ttran} \, \ldots \, \vect{p}_{|\mathcal{L}|}^{\Ttran}]^{\Ttran}$ and $\vect{p}_j = [p_{j1} \, \ldots \, p_{jK}]^{\Ttran}$, if $\check{\vect{g}}_{jk} = \vect{g}_{jk}^{\mathrm{scheme}}$ for all $j$ and $k$. Since the eigenvalues of $(\vect{I}_{K |\mathcal{L}|} - \vect{D} \boldsymbol{\Psi} )$ and $(\vect{I}_{K |\mathcal{L}|} - \vect{D} \boldsymbol{\Psi}^{\Ttran} )$ are the same, we can always select the DL powers according to \eqref{eq:DL-power}. It is straightforward to verify that $\vect{1}^{\Ttran} \vect{q}^{\star} = \vect{p}^{\Ttran} \vect{1}$, thus the total transmit power is the same in the DL and UL. Since the same SINRs as in the UL are achieved in the DL for any UE positions, the SE in \eqref{eq:asymptotic-SE-DL} follows directly from Theorems \ref{theorem-achievable-rate-conventional} and \ref{theorem-achievable-rate-pilotbased}.

\vspace{3mm}

\textbf{Proof of Lemma \ref{lemma:SE-impairment}:}
The derivations of Lemmas \ref{lemma:SE} and \ref{lemma:SE-DL} are based on the following principle: if we receive $s+n$, where $s$ is a Gaussian information signal and $n$ is uncorrelated noise, then an achievable SE is $\log_2 \big( 1 + \frac{|\mathbb{E}\{ s \}|^2}{ \mathbb{E}\{ |n|^2 \} + \mathbb{E}\{ |s|^2 \} - |\mathbb{E}\{ s \}|^2   } \big)$ \cite{Medard2000a}. For the hardware impairment models in \eqref{eq:system-model-impair} and \eqref{eq:system-model-DL-impair}, the received signals (after linear processing) behave as $\sqrt{1-\epsilon^2} (s + n) + \epsilon \eta$ instead, where $\mathbb{E}\{ |\eta |^2 \} = \mathbb{E}\{ |n|^2 \} + \mathbb{E}\{ |s|^2 \}$. Since the distortion $\eta$ is uncorrelated with $s$ and $n$ by assumption, $n+\eta$ is also uncorrelated with $s$ and the corresponding SINR is computed in \eqref{eq:lemma-impairment-derivation} at the top of the page. The only impact of the distortion is thus the $(1-\epsilon^2)$-factors in front of $|\mathbb{E}\{ s \}|^2$ in the numerator and denominator. The UL SINRs in \eqref{eq:SINR-value-impairment} follow directly from this observation, while the DL SINRs are achieved by also utilizing the power control policy from the proof of Theorem \ref{theorem:DLUL-duality}.

\bibliographystyle{IEEEtran}
\bibliography{IEEEabrv,refs}

\begin{IEEEbiography}{Emil Bj\"{o}rnson} (S'07, M'12) received his M.S. degree in Engineering Mathematics from Lund University, Sweden, in 2007. He received his Ph.D. degree in Telecommunications from the KTH Royal Institute of Technology, Stockholm, Sweden, in 2011. From 2012 to July 2014, he was a joint postdoc at Supélec, Gif-sur-Yvette, France, and at KTH Royal Institute of Technology. He is currently an Assistant Professor at the Department of Electrical Engineering (ISY) at Link\"oping University, Sweden.

His research interests include multi-antenna cellular communications, radio resource allocation, energy efficiency, massive MIMO, and network topology design. He is the first author of the textbook \emph{Optimal Resource Allocation in Coordinated Multi-Cell System} (Foundations and Trends in Communications and Information Theory, 2013). He is also dedicated to reproducible research and has made a large amount of simulation code publicly available. Dr. Bj\"{o}rnson 
 received the 2014 Outstanding Young Researcher Award from IEEE ComSoc EMEA and the 2015 Ingvar Carlsson Award. He has received 5 best paper awards for novel research on optimization and design of multi-cell multi-antenna communications: ICC 2015, WCNC 2014, SAM 2014, CAMSAP 2011, and WCSP 2009.
\end{IEEEbiography}

\begin{IEEEbiography}{Erik G. Larsson} received his Ph.D. degree from Uppsala University,
Sweden, in 2002.  Since 2007, he is Professor and Head of the Division
for Communication Systems in the Department of Electrical Engineering
(ISY) at Link\"oping University (LiU) in Link\"oping, Sweden. He has
previously been Associate Professor (Docent) at the Royal Institute of
Technology (KTH) in Stockholm, Sweden, and Assistant Professor at the
University of Florida and the George Washington University, USA.  In the spring of 2015 he was
a Visiting Fellow at Princeton University, USA, for four months.

His main professional interests are within the areas of wireless
communications and signal processing. He has published some 100 journal papers
on these topics, he is co-author of the textbook \emph{Space-Time
Block Coding for Wireless Communications} (Cambridge Univ. Press,
2003) and he holds 15 issued and many pending patents on wireless technology.

He has served as Associate Editor for several major journals, including the \emph{IEEE Transactions on
Communications} (2010-2014) and \emph{IEEE Transactions on Signal Processing} (2006-2010). 
He serves as  chair of the IEEE Signal Processing Society SPCOM technical committee in 2015--2016 and  
as chair of the steering committee for the \emph{IEEE Wireless Communications Letters} in 2014--2015.  He is the General Chair of the Asilomar Conference on Signals, Systems and Computers in 2015 (he was Technical Chair in
2012).  He received the \emph{IEEE Signal Processing Magazine} Best Column Award twice, in 2012 and 2014, and he is receiving the IEEE ComSoc Stephen O. Rice Prize in Communications Theory in 2015.
\end{IEEEbiography}

\begin{IEEEbiography}{M\'erouane Debbah} (SM'08, F'15)  entered the Ecole Normale Sup{\'e}rieure de Cachan (France) in 1996 where he received his M.Sc and Ph.D. degrees respectively. He worked for Motorola Labs (Saclay, France) from 1999-2002 and the Vienna Research Center for Telecommunications (Vienna, Austria) until 2003. From 2003 to 2007, he joined the Mobile Communications department of the Institut Eurecom (Sophia Antipolis, France) as an Assistant Professor. Since 2007, he is a Full Professor at CentraleSupelec (Gif-sur-Yvette, France). From 2007 to 2014, he was the director of the Alcatel-Lucent Chair on Flexible Radio. Since 2014, he is Vice-President of the Huawei France R\&D center and director of the Mathematical and Algorithmic Sciences Lab.

His research interests lie in fundamental mathematics, algorithms, statistics, information \& communication sciences research. He is an Associate Editor in Chief of the journal Random Matrix: Theory and Applications and was an associate and senior area editor for IEEE Transactions on Signal Processing respectively in 2011-2013 and 2013-2014. Dr. Debbah is a recipient of the ERC grant MORE (Advanced Mathematical Tools for Complex Network Engineering). He is a IEEE Fellow, a WWRF Fellow and a member of the academic senate of Paris-Saclay. He has managed 8 EU projects and more than 24 national and international projects. He received 14 best paper awards, among which the 2007 IEEE GLOBECOM best paper award, the Wi-Opt 2009 best paper award, the 2010 Newcom++ best paper award, the WUN CogCom Best Paper 2012 and 2013 Award, the 2014 WCNC best paper award, the 2015 ICC best paper award, the 2015 IEEE Communications Society Leonard G. Abraham Prize and 2015 IEEE Communications Society Fred W. Ellersick Prize as well as the Valuetools 2007, Valuetools 2008, CrownCom2009, Valuetools 2012 and SAM 2014 best student paper awards. He is the recipient of the Mario Boella award in 2005, the IEEE Glavieux Prize Award in 2011 and the Qualcomm Innovation Prize Award in 2012. He is the co-founder of the start-up Ximinds.
\end{IEEEbiography}

\end{document}